\documentclass[twocolumn]{aastex631}

\usepackage{mathtools}

\newcommand{\gaia}{\textit{Gaia}}
\newcommand{\rst}{\textit{Roman}}
\newcommand{\hst}{\textit{HST}}
\newcommand{\gaiahub}{\texttt{GaiaHub}}
\newcommand{\bpm}{\texttt{BP3M}}
\newcommand{\maspyr}{\mathrm{mas}~\mathrm{yr}^{-1}}
\newcommand{\masppixel}{\mathrm{mas}\cdot\mathrm{pixel}^{-1}}
\newcommand{\kmps}{\mathrm{km}~\mathrm{s}^{-1}}
\newcommand{\feh}{[\mathrm{Fe}/\mathrm{H}]}

\newcommand{\age}{{\mathrm{age}}}

\shorttitle{Bayesian Proper Motions with \textit{HST} and \textit{Gaia}}
\shortauthors{McKinnon et al.}

\received{October 30, 2023}
\revised{FUTURE DATE}
\accepted{FUTURE DATE}
\submitjournal{AAS Publishing}

\begin{document}

\title{\bpm: Bayesian Positions, Parallaxes, and Proper Motions derived from the \textit{Hubble~Space~Telescope} and \textit{Gaia} data}

\correspondingauthor{Kevin McKinnon}
\email{kevin.mckinnon@ucsc.edu}

\author[0000-0001-7494-5910]{Kevin A. McKinnon}
\affiliation{Department of Astronomy \& Astrophysics, University of California, Santa Cruz, 1156 High Street, Santa Cruz, CA 95064, USA}

\author[0000-0003-4922-5131]{Andr{\'e}s del Pino}
\affiliation{Centro de Estudios de F\'isica del Cosmos de Arag\'on (CEFCA), Unidad Asociada al CSIC, Plaza San Juan 1, E-44001, Teruel, Spain}

\author[0000-0002-6667-7028]{Constance M. Rockosi}
\affiliation{Department of Astronomy \& Astrophysics, University of California, Santa Cruz, 1156 High Street, Santa Cruz, CA 95064, USA}

\author[0009-0001-2827-1705]{Miranda Apfel}
\affiliation{Department of Astronomy \& Astrophysics, University of California, Santa Cruz, 1156 High Street, Santa Cruz, CA 95064, USA}

\author[0000-0001-8867-4234]{Puragra Guhathakurta}
\affiliation{Department of Astronomy \& Astrophysics, University of California, Santa Cruz, 1156 High Street, Santa Cruz, CA 95064, USA}

\author[0000-0001-7827-7825]{Roeland P. van der Marel}
\affiliation{Space Telescope Science Institute, 3700 San Martin Drive, Baltimore, MD 21218, USA}
\affiliation{Center for Astrophysical Sciences, The William H.~Miller III Department of Physics \& Astronomy, Johns Hopkins University, Baltimore, MD 21218, USA}

\author[0000-0001-8354-7279]{Paul Bennet}
\affiliation{Space Telescope Science Institute, 3700 San Martin Drive, Baltimore, MD 21218, USA}

\author[0000-0003-4207-3788]{Mark A.\ Fardal}
\affiliation{Eureka Scientific, 2452 Delmer Street, Suite 100, Oakland, CA 94602, U.S.A.}

\author[0000-0001-9673-7397]{Mattia Libralato}
\affiliation{AURA for the European Space Agency (ESA), ESA Office, Space Telescope Science Institute, 3700 San Martin Drive, Baltimore, MD 21218, USA}

\author[0000-0001-8368-0221]{Sangmo Tony Sohn}
\affiliation{Space Telescope Science Institute, 3700 San Martin Drive, Baltimore, MD 21218, USA}

\author[0000-0002-2732-9717]{Eduardo Vitral}
\affiliation{Space Telescope Science Institute, 3700 San Martin Drive, Baltimore, MD 21218, USA}

\author[0000-0002-1343-134X]{Laura L. Watkins}
\affiliation{AURA for the European Space Agency (ESA), ESA Office, Space Telescope Science Institute, 3700 San Martin Drive, Baltimore, MD 21218, USA}

\begin{abstract}
We present a hierarchical Bayesian pipeline, \texttt{BP3M}, that measures positions, parallaxes, and proper motions (PMs) for cross-matched sources between \textit{Hubble~Space~Telescope} (\textit{HST}) images and \textit{Gaia} -- even for sparse fields ($N_*<10$ per image) -- expanding from the recent \texttt{GaiaHub}\@ tool. This technique uses \textit{Gaia}-measured astrometry as priors to predict the locations of sources in \textit{HST} images, and is therefore able to put the \textit{HST} images onto a global reference frame without the use of background galaxies/QSOs. Testing our publicly-available code in the Fornax and Draco dSphs, we measure accurate PMs that are a median of 8-13 times more precise than \textit{Gaia} DR3 alone for $20.5<G<21~\mathrm{mag}$. We are able to explore the effect of observation strategies on \bpm\@ astrometry using synthetic data, finding an optimal strategy to improve parallax and position precision at no cost to the PM uncertainty. Using 1619 \textit{HST} images in the sparse COSMOS field (median 9 \gaia\@ sources per \hst\@ image), we measure \texttt{BP3M} PMs for 2640 unique sources in the $16<G<21.5~\mathrm{mag}$ range, 25\% of which have no \textit{Gaia} PMs; the median \bpm\@ PM uncertainty for $20.25<G<20.75~\mathrm{mag}$ sources is $0.44~\maspyr$ compared to $1.03~\maspyr$ from \gaia\@, while the median \bpm\@ PM uncertainty for sources without \gaia\@-measured PMs ($20.75<G<21.5~\mathrm{mag}$) is $1.16~\maspyr$. The statistics that underpin the \bpm\@ pipeline are a generalized way of combining position measurements from different images, epochs, and telescopes, which allows information to be shared between surveys and archives to achieve higher astrometric precision than that from each catalog alone.
\end{abstract}


\section{Introduction} \label{sec:intro}

High precision proper motions (PM) of individual stars have dramatically increased our understanding of kinematics in the Local Group. In particular, the \textit{Hubble~Space~Telescope} (\hst\@) has a rich history of providing the precise astrometry needed to make accurate motion measurements. This is thanks to detailed work over the last few decades to characterize and correct \hst\@ geometric distortions, describe point spread functions (PSFs), and define best practices for mapping images onto one another \citep[e.g.][]{Anderson_2004,Anderson_2006,Anderson_2007,Bellini_2009,Bellini_2011}. As a result, the PMs measured from multi-epoch \hst\@ imaging have been used to study key astronomical questions, such as the relative motion of the Milky Way (MW) and M31 \citep{Sohn_2012,vanderMarel_2012}, the mass of the MW \citep{Sohn_2018}, and MW stellar halo kinematic measurements in individual fields \citep{Cunningham_2019b}.

More recently, the \gaia\@ mission \citep{Gaia_2018} has facilitated substantial leaps in Local Group science by providing full astrometric solutions for millions of stars. In the MW, these results include the identification of our most massive merger, the Gaia-Sausage-Enceladus \citep[e.g][]{Helmi_2018,Belokurov_2018,Mackereth_2019}, detailed inventories of the progenitors that built the MW \citep[e.g.][]{Naidu_2020}, and nearly-complete catalogs of the motions of nearby globular clusters \citep{Vasiliev_2021}. 

While both of these telescopes have and will continue to fundamentally alter our knowledge about the local universe, they both have limitations. In \hst's case, the small field of view and relatively long observation times mean that only a small portion of the sky has been observed at multiple epochs with significant time baselines. For \gaia\@, their all-sky catalog necessarily does not see as faint as a standard \hst\@ observation, and the precision of their astrometric solutions decreases fairly significantly for their faintest magnitudes ($G\sim21~\mathrm{mag}$). \gaia's fairly bright limiting magnitude for stars with precise PMs restricts the spatial resolution on which it can measure average kinematics in the stellar halo. Consequently, many of the previous studies cited above focus on stellar populations over large portions of the sky when measuring the kinematics of the MW stellar halo, but a growing body of work has shown the benefit of being able to resolve (chemo)dynamics on small spatial scales \citep[e.g.][]{Cunningham_2019b,Iorio_2021,McKinnon_2023}. With better sky velocities, we can improve constraints on the formation and evolution history of our Galaxy \citep[e.g.][Apfel et al. in prep]{Cunningham_2022,Cunningham_2023}.

To reduce the effect of these limitations and to increase the astrometric constraining power of either telescope alone, recent studies have been exploring how to combine information between datasets from different telescopes. Specifically, using archival \hst\@ images that have long time baselines with \gaia\@, PMs that are factors of 10 to 20 times more precise than \gaia\@ alone have been measured for dwarf spheroidal galaxy and globular cluster stars \citep{Massari_2017,Massari_2018,Massari_2020,delPino_2022,Bennet_2022}, which have enabled internal velocity dispersion measurements. Techniques for cross-telescope combinations will become even more important as the field progresses further into the Big Data era of astronomy, especially as future missions come online (e.g. \textit{Nancy~Grace~Roman~Space~Telescope}, Rubin, Euclid).

Regardless of where the astrometry comes from, PMs in the MW can be more challenging to measure than PMs of distant sources. This is largely because the apparent motion on the sky can be quite large and the motion from parallax can become significant. There are three main ways that a star can appear to move between successive images: (1) apparent offset because of statistical uncertainty on position, (2) offsets due to parallax, and (3) offsets due to PM. In studies of distant sources, the motion from parallax can often be ignored. Because of relatively small \gaia\@ uncertainties for source positions, the impact that offsets between true positions and \gaia-measured positions have on PM measurements is usually small enough that it can also be ignored, especially as the time baseline increases. For extragalactic stars, then, all apparent motion is the result of PM, and the longer the time baseline between observations, the more precise and accurate the PM measurements. When the position uncertainties become large (e.g. for faint \gaia\@ sources) and the parallaxes become substantial (e.g. $D < 1$~kpc), then detailed and simultaneous accounting of all three motion components becomes necessary. One additional difficulty in measuring MW PMs, especially in studies of the stellar halo, comes from the fact that many fields are quite sparse; this limits the number of sources that can be used to constrain the transformation parameters that align images from multiple epochs and impacts the accuracy and precision of the resulting PMs. 

While the \gaiahub\@ pipeline of \citet{delPino_2022} was developed for and performs well in populated fields, a key motivation of this work was to develop a complementary pipeline that can also handle very sparse fields (e.g. $N_* < 10$). To address the challenges listed previously, we create a hierarchical Bayesian pipeline, named \bpm\@ (Bayesian Positions, Parallaxes, and Proper Motions), to simultaneously measure the positions, parallaxes, and PMs of all \gaia\@ sources in an \hst\@ image while also finding the best mapping of \hst\@ images onto \gaia\@. This package is publicly available on Github\footnote{\url{https://github.com/KevinMcK95/BayesianPMs}} and is designed to be used in combination with the \gaiahub\@ code. The underlying statistics are general in that they apply to any two or more sets of position measurements separated by time, regardless of the telescope they come from. In this way, it is also a useful tool for planning future observations.

In this paper, we describe a standard approach for measuring PMs in Section~\ref{sec:gaiahub}, which then motivates the statistics and pipeline, \bpm\@, we present in Section~\ref{sec:methods}. We examine and validate the pipeline's outputs in Section~\ref{sec:validation} using synthetic data. We then highlight some applications of \bpm\@ on real data in Section~\ref{sec:applications}, including sparse fields in the MW and nearby dwarf spheroidals. Finally, our complete set of results are summarized in Section~\ref{sec:summary}.

\section{Measuring Proper Motions} \label{sec:gaiahub}

To measure the motion of stars, one traditionally identifies the positions of sources at two epochs, measures the best transformation parameters between those two images, and transforms the coordinates of sources from one image into the other. The final offsets can then be used to measure the relative motion of each of the sources. When mapping one image onto another, a standard approach \citep{Anderson_2007} is to describe the position of a source in the first image, $(X,Y)$, in coordinates of the second image $(X',Y')$ using:
\begin{equation}
    \label{eq:tranformation_relation}
    \left(\begin{matrix}
            X'\\
            Y'\\
    \end{matrix}\right) = 
    \left(\begin{matrix}
            a&b\\
            c&d\\
    \end{matrix}\right) \cdot 
    \left[
    \left(\begin{matrix}
            X\\
            Y\\
    \end{matrix}\right) - 
    \left(\begin{matrix}
            X_0\\
            Y_0\\
    \end{matrix}\right)
    \right]+
    \left(\begin{matrix}
            W_0\\
            Z_0\\
    \end{matrix}\right)
\end{equation}
where the $(X_0,Y_0)$ and $(W_0,Z_0)$ vectors are the center of rotation in each coordinate system. The $(a,b,c,d)$ matrix accounts for differences in pixel scale, rotation, and skew as follows:
\begin{subequations}
    \label{eq:trans_term_definitions}
\begin{flalign}
        \mathrm{pixel~scale~ratio} &= \mathrm{PSR} = \sqrt{ad-bc}\ , \\
        \tan \theta &= \left(\frac{b-c}{a+d}\right)\ , \\
        \mathrm{on~axis~skew} &= \frac{1}{2}(a-d)\ , \\
        \mathrm{off~axis~skew} &= \frac{1}{2}(b+c)\ .
\end{flalign}
\end{subequations}

With this relationship, we see that there are 6 parameters that need to be fit when transforming one image onto another\footnote{We only need to fit for either $(X_0,Y_0)$ or $(W_0,Z_0)$ because we can fix one vector and the other will absorb those choices.}. In general, this would only require the positions of 3 sources found in both images, but in practice, many more sources are required to measure a robust transformation solution. 

To be clear, this technique measures the change in $(X,Y)$ as a function of time, but these measurements are relative to the population of sources in the image. For instance, a collection of stars moving with the same PM will only have a difference in translation at two epochs; fitting for transformation parameters will then yield zero change in the relative positions of all the stars. This is still a very useful technique in the cases where the relative motion of stars is of interest, such as for measuring internal kinematics of clusters and galaxies. 

To find the PM of the stars in an absolute reference frame, we require known anchor points that have no motion (e.g. background stationary sources like galaxies and QSOs). An alternative approach is to cross-match the stars in an image to an absolute-frame-calibrated survey to estimate the bulk velocity/average absolute velocity of the sources, which is the technique that the \gaiahub\@ pipeline \citep{delPino_2022} employs. In a high level overview, the \gaiahub\@ pipeline steps are as follows:
\begin{enumerate}
    \item Following previous techniques for \hst\@ data reduction \citep[e.g.][]{Bellini_2017,Bellini_2018,Libralato_2018,Libralato_2019,Libralato_2022}, fit \hst\@ images with different point spread functions (PSFs) to identify sources and measure their positions using \texttt{hst1pass} \citep{Anderson_2022};
    \item Cross-match \hst\@ sources with \gaia\@;
    \item Find the transformation parameters that minimize the offsets between the \hst\@ source positions and \gaia\@ positions for each image;
    \item With the best fit transformation parameters in hand, any remaining offsets are divided by the time baseline to give the relative PM for each source in an image;
    \item Use an average of \gaia\@-measured PMs to estimate the bulk velocity, giving PMs in an absolute frame;
    \item Combine the multiple \gaiahub\@ PM measurements of a source using an inverse-variance weighted average to get a final PM and uncertainty.
\end{enumerate}
As is shown in \citet{delPino_2022}, the pipeline performs well, especially in the medium density outskirts of nearby galaxies for which it was developed. Two key assumptions are required for \gaiahub\@ results to accurately describe a population: (1) the parallax motion of the sources are insignificant, and (2) there are enough sources such that their PM distribution is close to Gaussian. The second point is particularly important because the offsets between the sources in different images, which determines the transformation parameters, are minimized when fitting for the transformation. If there are a small number of sources, the individual offsets have an out-sized impact on the resulting transformation parameters; sources with high offsets between images (such as a fast-moving foreground star), can torque the transformation solution around to try to reduce what is an intrinsically large separation. Similarly, in the limiting  case where an image contains 3 sources, we can find transformation parameters such that there is no remaining offset, even if the sources truly have moved. In both cases, the transformation parameters are different from the truth, and this impacts the PM measurements of all sources in an image. \gaiahub\@ allows users to remove the impact of outlier PMs (e.g. foreground stars) by defining a co-moving population or to ``rewind'' sources using the \gaiahub-measures PMs while iterating on the transformation fitting, but both of these modes don't apply to halo populations that lack an obvious co-moving frame and have a small number of sources per \hst\@ image. 

Motivated by the PM improvements that \gaiahub\@ has shown when information from \gaia\@ and \hst\@ is combined together, we create a tool that enables \gaia+\hst\@ PM measurements in sparse fields to study the MW stellar halo. Because these sparse fields have low stellar density (e.g. $<20$ \gaia\@ sources per \hst\@ image) and can contain a significant fraction of nearby sources (e.g. foreground disk and nearby halo stars, which can have non-negligible parallax motion and relatively large sky velocities), our pipeline cannot use the same assumptions that go into \gaiahub\@. This realization is what ultimately informed our decision to model the motions and transformation parameters simultaneously, though we emphasize that the following pipeline benefits immensely from the \gaiahub\@ project. Specifically, the PSF fitting to measure centroids in \hst\@, the cross-matching between \hst\@ and \gaia\@, and the initial measurements of the transformation parameters from \gaiahub\@ are integral components of the pipeline we present in the remainder of this work.

\subsection{COSMOS Test Sample} \label{ssec:COSMOS_testbed}

To demonstrate the performance of \gaiahub\@ in sparse MW halo fields, we turn to the COSMOS field \citep{Nayyeri_2017,Muzzin_2013} from the Cosmic Assembly Near-infrared Deep Extragalactic Legacy Survey \citep[CANDELS;][PIs: S. Faber, H. Ferguson]{Grogin_2011,Koekemoer_2011}. COSMOS is located at a high Galactic latitude and near the Galactic anti-center ($l=236.8~\mathrm{deg}$, $b=42.1~\mathrm{deg}$) and boasts \hst\@ imaging that covers a large area of sky: $\sim2~\mathrm{deg}^{2}$ was imaged by ACS/WFC in 2003/2004, and 288~arcmin$^{2}$ of that area (i.e. $\sim4\%$) was imaged again in 2011/2012. From simulated COSMOS-like data -- presented in detail in Appendix~\ref{sec:generating_synthetic_data} -- PMs of halo and foreground disk stars in our magnitude range (i.e. $16<G<21.5~\mathrm{mag}$) can regularly be as large as $\sim 100~\maspyr$ in size.

\begin{figure}[t]
\begin{center}
\includegraphics[width=\linewidth]{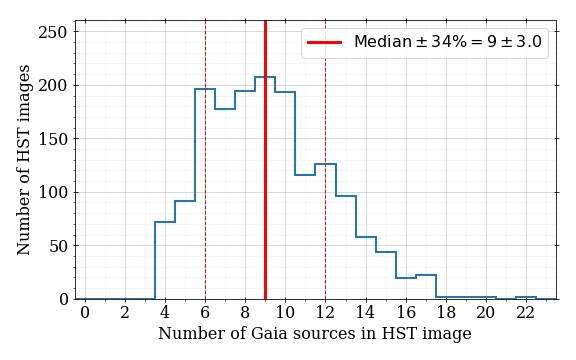}
\caption{Histogram of the number of \gaiahub\@ cross-matched sources between \gaia\@ and \hst\@ for the 1619 \hst\@ images in COSMOS within $0.5~\mathrm{deg}$ of the field's center. There is a median of 9 \gaia\@ sources in each \hst\@ frame, and a total of 2640 unique \gaia\@ sources.}
\label{fig:COSMOS_nstar_per_image}
\end{center}
\end{figure}

We analyse all COSMOS \hst\@ images within a $0.5~\mathrm{deg}$ radius of the field's center, which is $\sim39\%$ of the total COSMOS area. Figure~\ref{fig:COSMOS_nstar_per_image} shows a histogram of the number of sources matched between the \hst\@ frames and \gaia\@ for the 1619 \hst\@ images in our analysis. In these \hst\@ images, \gaiahub\@ finds 2640 unique \gaia\@ sources, with 4, 9, and 23 sources per image in the minimum, median, and maximum cases. To be clear, there are many repeat exposures at the same \hst\@ pointing as well as overlapping and multi-epoch regions of the COSMOS field, so all of the sources appear in more than one of the 1619 images. 

\begin{figure*}[t]
\begin{center}
\includegraphics[width=\linewidth]{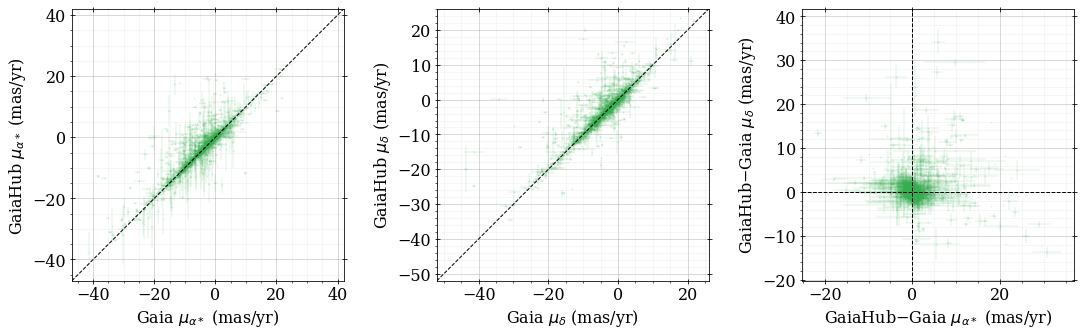}
\caption{Comparison of PMs measured by \gaia\@ and \gaiahub\@ for 1409 COSMOS sources. These sources are a subset of the 2640 unique COSMOS sources discussed in Figure~\ref{fig:COSMOS_nstar_per_image} (1985 of which have \gaia-measured PMs) that \gaiahub\@ determined to have well-measured \gaia\@ PMs for rewinding to \hst\@ epochs. The PMs generally fall along or agree with the one-to-one line (dashed black diagonals in the left and middle panels) within their uncertainties, though there are also many PM measurements that disagree significantly. This is largely because the \gaiahub\@ pipeline was not tuned for small numbers of sources per \hst\@ image or for the large PMs seen in the MW stellar halo.}
\label{fig:COSMOS_gaiahub_comp}
\end{center}
\end{figure*}

A comparison of the \gaia\@- and \gaiahub-measured PMs for 1264 COSMOS sources is presented in Figure~\ref{fig:COSMOS_gaiahub_comp}. The \gaiahub\@ analysis was run using the \texttt{--rewind\_stars} option, which iterates on the transformation fitting using \gaia\@ (in the first iteration) and \gaiahub\@ (in subsequent iterations) PMs to extract better solutions, especially when high-PM stars are present. However, the use of this option removes many sources that have poorly-measured \gaia\@ PMs, resulting in more than half of the sources being rejected by \gaiahub\@ in the COSMOS fields (i.e. 1659 used by \gaiahub\@ from the 2640 possible unique sources). While more relaxed \gaiahub\@ options produce PMs for all 2640 unique \gaia\@ sources in the COSMOS images, these results are sensitive to small-number statistics and poorly-measured \gaia\@ PMs that can shift the reference frame solution and yield extreme disagreements with \gaia.

In general, the \gaiahub\@ and \gaia\@ PMs follow the one-to-one line, and many PM measurements agree within their uncertainties. However, there are also some PMs that show significant disagreement between the two catalogs, as is evident in the rightmost panel. Using Mahalanobis distances between the PM measurements (i.e. accounting for their uncertainties) for sources with $<5\sigma$ disagreement between \gaia\@ and \gaiahub\@ PMs, we find that the \gaiahub\@ PM uncertainties would need to be inflated by a factor of $\sim1.7$ to explain the PM differences. For $\sim10\%$ these PMs, the \gaiahub\@ measurements are more than $5\sigma$ different from the corresponding \gaia\@ values, whereas a sample size of $\sim1000$ should expect to see no deviations this large.

Because \gaiahub\@ is not tuned for sparse fields, it is not surprising to see these disagreements. Some of this disparity comes from the fact that \gaiahub\@ is only able to propagate realistic uncertainties on the transformation parameters to the measured PMs when there are several repeat \hst\@ observations of the same field. For much of COSMOS, there are only $\sim4$ \hst\@ observations per field, which is not enough samples to obtain statistically robust propagation using \gaiahub\@'s approach. While the transformation uncertainty is usually a small component of the total error budget when there are a large number of sources, the transformation uncertainties in the sparse images of COSMOS are significant, so many of the \gaiahub\@ PM uncertainties are underestimated. Handling these sparse field cases therefore requires a new tool. 

\section{Bayesian Positions, Parallaxes, and Proper Motions: The \bpm~Pipeline} \label{sec:methods}

We build a hierarchical Bayesian tool, \bpm, that measures transformation parameters to map \hst\@ images onto \gaia\@ while also simultaneously measuring the PMs, parallaxes, and positions for the sources in an image. While it may appear computationally prohibitive to measure PMs, parallaxes, position of all stars in an image and the transformation parameters simultaneously, a few statistical tricks -- namely conjugacy between likelihood and prior distributions -- make this approach feasible. The pipeline is able to consider multiple \hst\@ images concurrently in a proper Bayesian fashion, which can significantly improve precision for the different motion components of each source.

We use the \gaia\@-measured positions, parallaxes, and PMs and corresponding uncertainties/covariance as prior distributions to describe each source position over time. This improves the measured astrometric solution because the priors allow the measured positions of the sources to be compared to their expected positions at the epoch of each image, rather than comparing the individual position measurements at each time. This approach is quite general in that it does not require identifying a clean kinematic sample or reference stars and background sources. In many cases, bright foreground stars with well-measured \gaia\@ astrometry can serve as anchors that help define the transformation solution. 

Throughout the statistics presented in this section and Appendix~\ref{sec:motion_statistics}, we refer to \hst\@ and \gaia\@ in the subscripts of the variables, but there is nothing about this math or its implementation in \bpm\@ that restricts us to only these telescopes. The statistical statements are generally true for any collection of measurements with at least two positions of the same source -- it applies to \hst\@+\hst\@, \hst\@+\rst\@, \hst\@+\gaia\@+\rst\@, etc -- though there are a minimum of three sources per image needed to be able to measure the 6 transformation parameters. For our COSMOS analysis, all \hst\@ images have 4 or more \gaia\@ sources.  

\begin{deluxetable*}{cc}
\tablecaption{Definitions of fitting parameters. In general, $H$ refers to an \hst\@ value, $G$ refers to an \gaia\@ value, an apostrophe indicates a prior measurement, $j$ refers to the \hst\@ image number, and $i$ refers to the source index.
\label{tab:parameter_definitions}}
\tablehead{
\colhead{Parameter} & \colhead{Description}} 
\startdata
$\left(a_j,b_j,c_j,d_j \right)$ & Transformation matrix parameters for the $j$-th \hst\@ frame to the $j$-th pseudo \gaia\@ frame\\
$(X_{0,j},Y_{0,j})$ & Center coordinate in the $j$-th \hst\@ frame\\
$(W_{0,j},Z_{0,j})$ & Center coordinate of $j$-th pseudo \gaia\@ frame\\
$(X_{H,i,j}',Y_{H,i,j}')$ & Observed coordinate of $i$-th source in the $j$-th \hst\@ frame\\
$\sigma_{H,i,j}$ & \gaiahub-measured pixel position uncertainty (in both x and y) of the $i$-th source in the $j$-th \hst\@ frame \\
$(X_{G,i,j}',Y_{G,i,j}')$ & Observed coordinate of $i$-th source in the $j$-th pseudo \gaia\@ frame\\
$\mathrm{PS}_{G,j}$ & Pseudo pixel scale of $j$-th pseudo \gaia\@ frame\\
$\vec \theta_i'^T = (\alpha_i',\delta_i')$ & \gaia-measured position vector for the $i$-th source\\
$(\sigma_{\alpha*,i}',\sigma_{\delta,i}',\rho_{\alpha*,\delta,i}')$ & \gaia-measured position uncertainties and correlation for the $i$-th source\\
$\mathrm{plx}'_i$ & \gaia-measured parallax for the $i$-th source\\
$\sigma_{\mathrm{plx},i}'$ & \gaia-measured parallax uncertainty for the $i$-th source\\
$\vec \mu_i'^T = (\mu_{\alpha*,i}',\mu_{\delta,i}')$ & \gaia-measured PM vector for the $i$-th source\\
$(\sigma_{\mu_{\alpha*},i}',\sigma_{\mu_{\delta},i}',\rho_{\mu_{\alpha*},\mu_{\delta},i}')$ & \gaia-measured PM uncertainties and correlation for the $i$-th source\\
$\vec{\Delta \theta}^T = (\Delta \alpha*_i,\Delta \delta_i)$ & True position offset vector for the $i$-th source\\
$\mathrm{plx}_i$ & True parallax for the $i$-th source\\
$\vec \mu_i'^T = (\mu_{\alpha*,i},\mu_{\delta,i})$ & True PM vector for the $i$-th source\\
$t_{G} = \mathrm{J2016.0}$ & Time of \gaia\@ frame observation\\
$t_{H,j}$ & Time of $j$-th \hst\@ frame observation\\
$\Delta t_j = t_{H,j}-t_{G}$ & Time between \gaia\@ and $j$-th \hst\@ frame (negative for older \hst\@ images)\\
$\vec{\Delta \mathrm{plx}}_{i,j} $ & Offset-per-parallax vector for $i$-th source between \gaia\@ and $j$-th \hst\@ frame\\
\enddata
\end{deluxetable*}

\subsection{The Statistics} \label{ssec:main_statistics}

This subsection will be quite technical in detail, so readers who are less interested in the formal statistics are encouraged to skip to Section~\ref{ssec:pipeline_description}. 

First, we define many of the key variables and parameters in Table~\ref{tab:parameter_definitions}. While we are ultimately concerned with the PMs in RA and Dec, it is convenient to work in a 2D plane projection (e.g. what a detector sees) when comparing images. Because \gaia's measurements do not correspond to positions in a true image, we follow \gaiahub's approach of transforming the \gaia\@ coordinates into XY coordinates on a pseudo image using a tangent-plane projection
\begin{subequations}
    \label{eq:radec_to_xy}
\begin{flalign}
        r_{G,i,j} &= \sin\left(\delta_{G,j,0}\right)\sin\left(\delta_{G,i}'\right)\\
        &+\cos\left(\delta_{G,j,0}\right)\cos\left(\delta_{G,j}'\right)\cos\left(\alpha_{G,i}'-\alpha_{G,j,0}\right) \nonumber \ , \\ \nonumber \\
        \mathrm{rad2pix}_{G,j} &= \frac{180}{\pi}\cdot 3600 \cdot 1000\cdot\mathrm{PS}_{G,j}^{-1}~\mathrm{\frac{pixels}{radian}} \ , \\ \nonumber \\
        X_{G,i,j} &= X_{G,j,0} - \mathrm{rad2pix}_{G,j}\\
        &\times \frac{\cos\left(\delta_{G,i}'\right) \sin\left(\alpha_{G,i}'-\alpha_{G,j,0}\right)}{r} \ , \nonumber \\ \nonumber \\
        Y_{G,i,j} &= Y_{G,j,0} \\
        &+ \mathrm{rad2pix}_{G,j}~\left[\frac{\cos\left(\delta_{G,j,0}\right)\sin\left(\delta_{G,i}'\right)}{r}\right. \nonumber \\
        &\left.-\frac{\sin\left(\delta_{G,j,0}\right)\cos\left(\delta_{G,i}'\right)\cos\left(\alpha_{G,i}'-\alpha_{G,j,0}\right)}{r}\right] \ , \nonumber
\end{flalign}
\end{subequations}
where the $(\alpha_{G,i}',\delta_{G,i}')$ are the \gaia-measured coordinates of source $i$ in radians, $\left(\alpha_{G,j,0},\delta_{G,j,0}\right)$ are the coordinates of the center of the \gaia\@ pseudo image when considering \hst\@ image $j$, $\left(X_{G,j,0},Y_{G,j,0}\right)$ are the pixel coordinates of the center of the \gaia\@ pseudo image, and $(X_{G,i,j},Y_{G,i,j})$ are the coordinates of source $i$ in the \gaia\@ pseudo image in pixels. The $\left(\alpha_{G,j,0},\delta_{G,j,0}\right)$ and $\left(X_{G,j,0},Y_{G,j,0}\right)$ are inherited from the \gaiahub\@-measured values for each \hst\@ image, which come from an average of the source positions, and these values are not left as free parameters during the fitting process. The chosen pixel scale $\mathrm{PS}_{G,j}$ for each \gaia\@ pseudo image is set to the nominal \hst\@ pixel scale that it is paired with (e.g. $50~\masppixel$ for \hst's ACS/WFC). Then, changes in the RA and Dec can be transformed to changes in XY using the Jacobian matrix:
$$\pmb J_{i,j} = \left(\begin{matrix}
\frac{\delta X_{G,i,j}}{\delta \alpha*_{G,i}'}&\frac{\delta X_{G,i,j}}{\delta \delta_{G,i}'}\\
\frac{\delta Y_{G,i,j}}{\delta \alpha*_{G,i}'}&\frac{\delta Y_{G,i,j}}{\delta \delta_{G,i}'}\\
\end{matrix}\right)$$
which is usually very close to
$$\pmb J_{j} = \frac{1}{\mathrm{PS_{G,j}}}\left(\begin{matrix}
-1&0\\
0&1\\
\end{matrix}\right)$$
when $(\Delta\alpha*,\Delta\delta)$ are in mas and $(\Delta X_{G,i,j},\Delta Y_{G,i,j})$ are \gaia\@ pseudo pixels. We note that this simplification only holds for small field-of-view detectors. In the cases of large PMs and large time baselines, the off-diagonal elements can start to become important, so we opt to use the more general version in our pipeline.

To set up the probability model, there are a few remaining key terms that we need to define:
$$\pmb R_j = \left(\begin{matrix}
a_j&b_j\\
c_j&d_j\\
\end{matrix}\right)$$
which is the transformation matrix for the $j$-th \hst\@ frame to the $j$-th pseudo \gaia\@ frame,
$$\pmb V_{H,i,j} = \left(\begin{matrix}
\sigma_{H,i,j}^2 & 0\\
0 & \sigma_{H,i,j}^2\\
\end{matrix}\right)$$
which is the \gaiahub-measured pixel position covariance matrix for the $i$-th source in the $j$-th \hst\@ frame,
$$\pmb V_{\theta,i} = \left(\begin{matrix}
\sigma_{\alpha*,i}'^2 & \rho_{\alpha*,\delta,i}'\cdot \sigma_{\alpha*,i}' \cdot \sigma_{\delta,i}'\\
\rho_{\alpha*,\delta,i}' \cdot \sigma_{\alpha*,i}'\cdot \sigma_{\delta,i}' & \sigma_{\delta,i}'^2\\
\end{matrix}\right)$$
which is the \gaia-measured position covariance matrix for the $i$-th source, and finally
$$\pmb V_{\mu,i} = \left(\begin{matrix}
\sigma_{\mu_{\alpha*},i}'^2 & \rho_{\mu_{\alpha*},\mu_{\delta},i}'\cdot \sigma_{\mu_{\alpha*},i}'\cdot \sigma_{\mu_{\delta},i}'\\
\rho_{\mu_{\alpha*},\mu_{\delta},i}' \cdot \sigma_{\mu_{\alpha*},i}'\cdot \sigma_{\mu_{\delta},i}' & \sigma_{\mu_{\delta},i}'^2\\
\end{matrix}\right)$$
which is the \gaia-measured PM covariance matrix for the $i$-th source. In these equations and the ones that will follow, our convention is to show matrices (and matrices of matrices) using a bold-faced typesetting.

The fundamental relationships between the prior \gaia\@ measurements and the true parameter for source $i$ are given by:
\begin{subequations}
\begin{flalign}
        \vec \theta_i' &= \vec\theta_i + \vec{\Delta\theta}_i\ , \\
        p(\vec{\Delta\theta}_i' | \vec{\Delta\theta}_i) &= \mathcal{N}\left(\vec{\Delta\theta}_i' | \vec{\Delta\theta}_i, \pmb V_{\theta,i} \right)\ , \\
        p(\mathrm{plx}_i'|\mathrm{plx}_i) &= \mathcal{N}\left(\mathrm{plx}_i' | \mathrm{plx}_i, \sigma'_{\mathrm{plx},i} \right)\ , \\
        p(\vec \mu_i' | \vec \mu_i) &= \mathcal{N}\left(\vec \mu_i'|\vec \mu_i, \pmb V_{\mu,i} \right)\ ,
\end{flalign}
\end{subequations}
which says that the \gaia\@-measured values are offset from the true values by noise dictated by their \gaia\@-measured uncertainties. While we have explicitly included a mean for the \gaia\@-measured position offset vector $\vec{\Delta \theta}_i'$ to be as general as possible, in practice $\vec{\Delta \theta}_i' = \vec 0$ because \gaia\@ has no expected offset from the positions it reports. In the form of the probability statements above, we have assumed that there is no correlation between, for example, parallax and $\mu_{\alpha*}$, which is not exactly correct because \gaia\@ has measurements for these correlations. Our choices, however, make the following math easier, though it comes at the cost of some additional constraining power from the \gaia\@ priors being ignored. The \gaia-measured correlations between these parameters are indeed included in future sections of this work when we compare the \bpm\@ distributions to \gaia's results. Future versions of the \bpm\@ pipeline will work to incorporate these prior correlations, which will lead to even tighter posterior distributions on position, parallax, and PMs. 

While it may seem nonphysical, \gaia\@ observed parallax measurements can be negative \citep{Lindegren_2018}, but it is important to remember that the observed parallax values define the mean of a distribution whose width/uncertainty usually places a large amount of probability in positive parallaxes; in this way, we must treat the \gaia\@ astrometric measurements as distributions and not individual points. \citet{Luri_2018}, for example, explain how the definitions of motion on the sky (both by \gaia\@ as well as in this work) technically allow for negative observed parallax values, which is especially likely to occur when the position uncertainty is relatively large compared to the size of the parallax motion (e.g. see their Section 3 and Figure 2); as a result, they remind the reader that the \gaia\@ observed parallaxes should not be thought of as a direct measurement of distance, and instead, distances need to be estimated by proper statistical modelling of the information contained in the astrometric solution distributions. 

Because some of the sources in an \hst\@ image have no \gaia\@-measured parallaxes or PMs, we find it useful to put a population/global prior on the PMs and parallaxes:
\begin{subequations}
\begin{flalign}
        p(\mathrm{plx}_i | \hat{\mathrm{plx}}, \sigma_{\hat{\mathrm{plx}}} ) &= \mathcal{N}\left(\mathrm{plx}_i | \hat{\mathrm{plx}}, \sigma_{\hat{\mathrm{plx}}} \right)\ , \\
        p(\vec \mu_i | \hat{\mu}, \pmb V_{\hat{\mu}}) &= \mathcal{N}\left(\vec \mu_i | \hat{\mu}, \pmb V_{\hat{\mu}} \right)\ ,
\end{flalign}
\end{subequations}
which says that there are some global distributions that the true parallaxes and PMs of the sources originate from. While we are free to play with the parameters of the population distributions, we note that they do need to be Gaussian in form so that we retain the necessary conjugacy. We choose to use diffuse hyperpriors, with the goal of minimally impacting the sources with \gaia-measured parallaxes and PMs while offering some guidance to the sources without. In the future, the parallax global prior could be made more constraining in cases where the distances are better known (e.g. clean populations of extra-Galactic stars) or using a magnitude-dependent parallax prior to better incorporate our understanding of the distribution of stars in the MW. Similar changes could also be made to the PM global prior. For the current version of the pipeline, which focuses on faint stars in the MW stellar halo and extra-Galactic sources, we choose the parallax prior to have a mean of 0.5~mas and width of 10.0~mas; when generating synthetic MW thick disk and halo stars (described in Appendix~\ref{sec:generating_synthetic_data}), we find that these choices contain 99.9\% of the synthetic parallaxes within $1\sigma$ and 100\% of the synthetic parallaxes within $3\sigma$ (maximum parallax of $\sim28$~mas). For the PM prior, we use the sources with \gaia\@-measured PMs to estimate a mean and covariance matrix, and then we multiply that covariance matrix by a factor of $10^2$ to guard against the possibility that some of the sources without \gaia\@ PMs are significantly different in their PM from the other sources.  

The information linking \gaia\@ to \hst\@ image $j$ for source $i$ is then given by:
\begin{subequations} \label{eq:motion_equations}
\begin{flalign}
        \vec{\Delta d}_{G,i,j} &= \left(\begin{matrix}X_{G,i,j}\\Y_{G,i,j}\\\end{matrix}\right)-\pmb R_j\cdot \left(\begin{matrix}X_{H,i,j}-X_{0,j}\\Y_{H,i,j}-Y_{0,j}\\\end{matrix}\right)-\left(\begin{matrix}W_{0,j}\\Z_{0,j}\\\end{matrix}\right)\ , \\
        \vec{\Delta m}_{i,j} &= \vec\mu_i\cdot \Delta t_j + \mathrm{plx_i}\cdot\vec{\Delta\mathrm{plx}}_{i,j} - \vec{\Delta\theta}_i \ , \\
        \vec{\Delta d}_{G,i,j} &\sim \mathcal{N}\left( \pmb J_{i,j}\cdot \vec{\Delta m}_{i,j},\pmb V_{d,i,j} = \pmb J_{i,j}\cdot \pmb V_{\theta,i}\cdot \pmb J_{i,j}^T\right. \nonumber \\
        &\left. \hspace{4cm}+\pmb R_{j}\cdot \pmb V_{H,i,j}\cdot \pmb R_{j}^T\right)\ ,
\end{flalign}
\end{subequations}
which says that the measured offset between the \gaia\@ and \hst\@ positions (after applying the transformation) in the pseudo \gaia\@ image is distributed around the offset implied by the sum of the motion from PM, parallax, and uncertainty in position. As mentioned back in the description of Equation~\ref{eq:tranformation_relation}, we are able to hold $(X_{0,j},Y_{0,j})$ fixed as constants while the $(W_{0,j},Z_{0,j})$ are left as free parameters during the fitting. We note that the relationship in Equation~\ref{eq:motion_equations} ignores the impact that radial motion has on changing the distance to a star between observations \citep[e.g. see see Section 2.3.1 of][]{van_de_Ven_2006} -- and therefore the magnitude of the PM and parallax -- though this is a safe assumption because the distance change is almost always extremely small (e.g. a star with LOS velocity of $1000~\kmps$ would only experience a distance change of 0.01~pc after 10 years). 

\begin{figure}[ht]
\begin{center}
\includegraphics[width=\linewidth,height=6in,keepaspectratio, trim={0cm 0cm 0cm 0cm},clip]{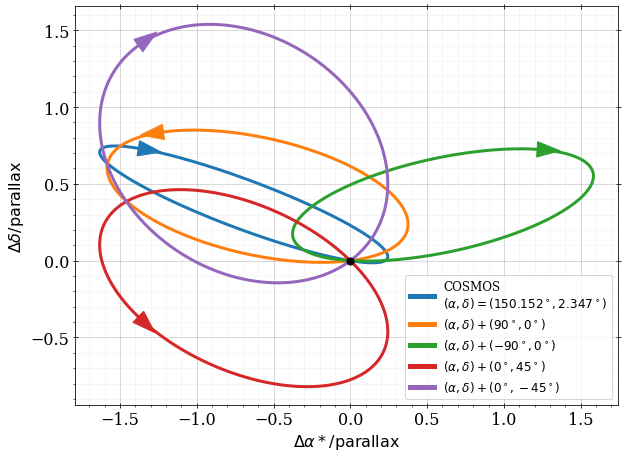}
\caption{Examples of parallax in the plane of the sky for different fields. The origin (black point) is the time that corresponds to \gaia\@ observations (i.e. J2016.0), and each ellipse is the path that a star with a parallax of $1~\mathrm{mas}$ sweeps out on the sky over the course of a year, with the direction of motion shown by the arrowhead. These ellipses demonstrate why the ability to constrain precise parallaxes is highly dependent on the exact observation times as well as position on the sky.}
\label{fig:parallax_motion}
\end{center}
\end{figure}

The $\vec{\Delta \mathrm{plx}}_{i,j}$ term, which we refer to as the parallax offset vector, is a 2D vector that defines the direction and magnitude of the offset between two observations as a result of parallax motion for a source with a parallax of 1~mas; in this way, the parallax offset vector can be multiplied by any parallax value (i.e. $\mathrm{plx}_i\cdot \vec{\Delta \mathrm{plx}}_{i,j}$) to find the appropriate offset in $(\Delta \alpha*,\Delta\delta)$ between two observations as a result of parallax motion. To measure the $\vec{\Delta \mathrm{plx}}_{i,j}$, we use the \hst\@ and \gaia\@ observation times, $(t_{H,j},t_{G})$, the position of the source in \gaia\@, $(\alpha_i',\delta_i')$, and built-in functions of \texttt{astropy} \citep{astropy_2013,astropy_2018,astropy_2022}. Examples of the parallax motion for different positions on the sky are shown in Figure~\ref{fig:parallax_motion}, where the \gaia\@ observation time (i.e. J2016.0) is at the origin and the orbits trace out the parallax motion over the course of a year; the parallax offset vector at any time is simply the vector that connects the corresponding point on the ellipse to the origin. 

With the above definitions in hand, we can use the distributions on the $\vec{\Delta d}_{G,i,j}$ vectors for all the sources in all images we are considering to construct the following likelihood distribution:
\begin{equation}
    \begin{split}
        p(&\pmb{X_{H}}, \pmb{Y_{H}}, \pmb{X_{G}}, \pmb{Y_{G}} | \vec a,\vec b,\vec c,\vec d,\vec{W_{0}},\vec Z_{0},\pmb \mu,\pmb{\mathrm{plx}},\pmb{\Delta \theta}) \\
        & = \prod_{i=1}^{n_{*}} \prod_{j=1}^{n_{im}}\mathcal{N}\left( \vec{\Delta d_{G,i,j}} | \pmb J_{i,j} \cdot \vec{\Delta m}_{i,j},\pmb V_{d,i,j} \right).
    \end{split}
\end{equation}
We note here that we have purposefully chosen to omit writing out the dependency on a few of the explanatory variables (e.g. $\Delta t_j$, the covariance matrices) for ease of reading the math. 

From Bayes' Law, we arrive at the following posterior:
\begin{equation} \label{eq:full_posterior}
    \begin{split}
        p&\left(\vec a,\vec b,\vec c,\vec d,\vec{W_{0}},\vec Z_{0},\pmb \mu,\vec{\mathrm{plx}},\pmb{\Delta \theta} | \right. \\
        &\hspace{2cm} \left. \pmb{X_{H}}, \pmb{Y_{H}}, \pmb{X_{G}}, \pmb{Y_{G}}, \hat{\mathrm{plx}},\sigma_{\hat{\mathrm{plx}}}, \hat{\mu}, \pmb V_{\hat\mu} \right) \\
        \propto& \left[\prod_{j=1}^{n_{im}} p(a_j,b_j,c_j,d_j,W_{0,j},Z_{0,j})\right] \cdot \\
        & \prod_{i=1}^{n_*} \left\{ p(\vec \mu_i' | \vec \mu_i) \cdot p(\vec \mu_i | \hat \mu, \pmb V_{\hat\mu}) \cdot p(\mathrm{plx}_i'|\mathrm{plx}_i) \cdot p(\mathrm{plx}_i| \hat{\mathrm{plx}}) \cdot \right.\\
        & \hspace{0.65cm} \left. p(\vec{\Delta \theta}_i' | \vec{\Delta\theta}_i) \cdot \prod_{j=1}^{n_{im}} \mathcal{N}\left( \vec{\Delta d_{G,i,j}} | \pmb J_{i,j} \cdot \vec{\Delta m}_{i,j},\pmb V_{d,i,j} \right) \right\}
    \end{split}
\end{equation}
where $p(a_j,b_j,c_j,d_j,W_{0,j},Z_{0,j})$ is the prior distribution on the transformation parameters and the remaining distributions have been defined above. In many of our initial analyses, we chose to use a flat prior on the transformation parameters, which yielded reasonable results. However, we soon realized from our analysis of a few hundred \hst\@ ACS/WFC images that there are relatively tight constraints for our expectations of the pixel scale ratio and skew terms. To that end, when using \hst\@ ACS/WFC data, we choose to employ a few relatively diffuse/not-overly-constraining priors on the skews, pixel scale ratio, and \hst\@ angle, as well as the $(W_{0,j},Z_{0,j})$ vector:
\begin{subequations}
\begin{flalign}
        p(\mathrm{skew}_j &| \mathrm{\hst~ACS/WFC}) \hspace{-2cm} \\
        &= \mathcal{N}\left(\mathrm{skew}_j | \mu=\mathrm{skew}_j', \sigma=5\times10^{-4} \right) \ , \nonumber\\
        p(\mathrm{PSR}_j &| \mathrm{\hst~ACS/WFC}) \hspace{-2cm} \\
        &= \mathcal{N}\left(\mathrm{PSR}_j | \mu=\mathrm{PSR}_j', \sigma=5\times10^{-4} \right) \ , \nonumber\\
        p(\theta_j) &= \mathcal{N}\left(\theta_j | \mu=\theta_j', \sigma=1^\circ \right) \ , \\
        p(W_{0,j}) &= \mathcal{N}\left(W_{0,j} | \mu=W_{0,j}', \sigma=10~\mathrm{pixels} \right) \ , \\
        p(Z_{0,j}) &= \mathcal{N}\left(Z_{0,j} | \mu=Z_{0,j}', \sigma=10~\mathrm{pixels} \right)\ ,
\end{flalign}
\end{subequations}
where we use the same prior on both the on- and off-axis skew terms. The prior means of each distribution come from the parameters measured by the previous iteration of transformation parameter fitting; in the case of the first iteration, we use the \gaiahub\@ outputs. In this way, we tell the pipeline that the transformation solution is likely nearby the previous iteration's solution, while the relatively large widths allows the values to change significantly if necessary. To transform from priors in angle, skews, and PSR, we need to account for a transformation Jacobian:
$$\pmb J_{\mathrm{TR,j}} = \left(\begin{matrix}
    \frac{\delta~\mathrm{on~skew}_j}{\delta a_j} & \frac{\delta ~\mathrm{off~skew}_j}{\delta a_j} & \frac{\delta \mathrm{PSR}_j}{\delta a_j} & \frac{\delta \theta_j}{\delta a_j}\\
    \frac{\delta ~\mathrm{on~skew}_j}{\delta b_j} & \frac{\delta ~\mathrm{off~skew}_j}{\delta b_j} & \frac{\delta \mathrm{PSR}_j}{\delta b_j} & \frac{\delta \theta_j}{\delta b_j}\\
    \frac{\delta ~\mathrm{on~skew}_j}{\delta c_j} & \frac{\delta ~\mathrm{off~skew}_j}{\delta c_j} & \frac{\delta \mathrm{PSR}_j}{\delta c_j} & \frac{\delta \theta_j}{\delta c_j}\\
    \frac{\delta ~\mathrm{on~skew}_j}{\delta d_j} & \frac{\delta ~\mathrm{off~skew}_j}{\delta d_j} & \frac{\delta \mathrm{PSR}_j}{\delta d_j} & \frac{\delta \theta_j}{\delta d_j}\\
\end{matrix}\right)$$
such that
$$p(a_j,b_j,c_j,d_j) = p(\mathrm{on~skew}_j,\mathrm{off~skew}_j,\mathrm{PSR}_j,\theta_j) \cdot |\pmb J_{\mathrm{TR,j}} |.$$ It should be noted that these priors may not always apply to all sets of ACS/WFC data, so this may need to be adjusted for other fields/applications. 

The complete forms of the posterior conditional distributions for each motion component or source $i$ are listed in Appendix~\ref{sec:motion_statistics}, defining the following important distributions:
\begin{subequations}
\begin{flalign}
        &p(\mathrm{plx}_i | \vec a,\vec b,\vec c,\vec d,\vec{W_{0}},\vec Z_{0}, \dots) \ , \\
        &p(\vec \mu_i | \mathrm{plx}_i, \vec a,\vec b,\vec c,\vec d,\vec{W_{0}},\vec Z_{0}, \dots) \ , \\
        &p(\vec{\Delta \theta_i} | \vec \mu_i, \mathrm{plx}_i, \vec a,\vec b,\vec c,\vec d,\vec{W_{0}},\vec Z_{0}, \dots)\ ,
\end{flalign}
\end{subequations}
where the ellipsis refers to the other variables that each distribution depends on. One key takeaway from the statistics that went into constructing the results in Appendix~\ref{sec:motion_statistics} is that the conjugacy between all distributions (i.e. all distributions are normally distributed/Gaussians) allows us to fairly easily combine multiple distributions to arrive at well-defined and closed-form posterior distributions. 

With these posterior conditionals in hand, we can perform the following steps for each source for a given set of transformation parameters:
\begin{enumerate}
    \item Draw samples of $\mathrm{plx}_i$ from $p(\mathrm{plx}_i |  \dots)$;
    \item Use those $\mathrm{plx}_i$ samples to draw samples of $\vec \mu_i$ from $p(\vec \mu_i | \mathrm{plx}_i, \dots)$;
    \item Use those $\left( \vec \mu_i, \mathrm{plx}_i \right)$ samples to draw $\vec{\Delta \theta_i}$ samples from $p(\vec{\Delta \theta_i} | \vec \mu_i, \mathrm{plx}_i, \dots)$.
\end{enumerate}
Once we have samples of the PMs, parallaxes, and position offsets for each source, we can calculate the posterior probability of a set of transformation parameters. We do this by marginalizing over the individual samples of the PMs, parallaxes, and position offsets using Bayes' Law such that
    \begin{equation}
        \begin{split}
            p(&\vec a,\vec b,\vec c,\vec d,\vec{W_{0}},\vec Z_{0} | \dots) \\
            &= \frac{p(\vec a,\vec b,\vec c,\vec d,\vec{W_{0}},\vec Z_{0},\pmb \mu,\vec{\mathrm{plx}},\pmb{\Delta \theta} | \dots)}{\prod_{i=1}^{n_{*}} p(\vec \mu_i, \mathrm{plx}_i, \vec{\Delta \theta_i} | \vec a,\vec b,\vec c,\vec d,\vec{W_{0}},\vec Z_{0},\dots)}.
        \end{split}
    \end{equation}
which is independent of the particular values of the PMs, parallaxes, and positions of each source. This relationship is the key that makes the simultaneous fitting of the 6 transformation parameters per image and 5 motion parameters per source feasible; because we can quickly draw the PM, parallax, and position samples directly from a known posterior distribution given a set of transformation parameters, we can efficiently calculate the posterior probability of the transformation parameters alone. Because we can calculate posterior probabilities for a given set of transformation parameters, we are able to sample those parameters from the posterior distribution using a Metropolis-Hastings (MH) MCMC algorithm. Then, for each sample of transformation parameters, we can sample from the posterior conditional distributions of parallaxes, PMs, and position offsets for each source. In this way, the uncertainty of the transformation fitting is propagated to the resulting PM, parallaxes, and positions. 

\subsection{The Pipeline} \label{ssec:pipeline_description}

In terms of implementing the statistics in the \bpm\@ pipeline, the general steps are as follows:
\begin{enumerate}
    \item Read in the position, parallax, PM data from \gaia\@, where it exists, and the corresponding positions in the \hst\@ frames from the \gaiahub\@ output catalogs\footnote{To be clear, \gaiahub\@ analyses the \hst\@ images, and \bpm\@ operates on the outputs of the \hst\@ image analysis from \gaiahub\@. \bpm\@ does not interact with the raw \hst\@ images, though we will use ``\hst\@ images'' as a synonym for ``\gaiahub-produced \hst\@ analysis catalog'' throughout the text of this work.};
    \item Use the \gaiahub\@ transformation parameter values as starting guesses;
    \item Using sources with \gaia\@ priors on PM, estimate the global PM distribution;
    \item Draws samples for the transformation parameters using MH-MCMC, evaluating posterior probabilities using only sources with \gaia\@ priors when measuring the transformation;
    \item Given a set of transformation parameters, draw samples of positions, parallaxes, and PMs for all sources;
    \item Identify outliers (e.g. bad cross-matches between \gaia\@ and \hst\@) as sources that have more than $2\sigma$ disagreement between the expected and observed positions in each \hst\@ image;
    \item Re-estimate the global PM distribution using the new posteriors;
    \item Repeat the fitting process using the non-outlier sources, including sources without \gaia\@ priors;
    \item Compare the new list of outliers to a union of the new list and the most recent previous list of outliers. If the new list length is more than 10\% different from the length of the union, repeat the fitting with the new list of outliers, and stop otherwise.
\end{enumerate}
This process requires a minimum of two iterations to achieve good results, though it isn't uncommon for the outlier list to change enough that a third iteration is required. In terms of processing time, running \bpm\@ on a 2016 MacBook Pro for an \hst\@ image with $\sim200$ sources takes approximate 15~minutes to complete its analysis, while an image with $\sim10$ sources can be analysed in around 5~minutes. Increasing the number of \hst\@ images being analysed together effectively multiplies the computation time by the number of images; this is largely because we need to fit 6 additional transformation parameters for each additional image, and necessarily need to increase the number of MCMC walkers and number of MCMC iterations. Best practice is to analyse all images independently before combining them to save time on searching the transformation parameter space. 

Because the \hst\@ images are mapped onto the global reference frame of \gaia\@, the transformation parameters that \bpm\@ measures are also useful in constraining non-\gaia\@ sources. Once the best set of transformation parameters between \hst\@ and \gaia\@ have been measured, we can return to the list of all sources in the \hst\@ image as measured by \gaiahub\@, which can include sources much fainter than \gaia\@ can see (i.e. $G>21.5~\mathrm{mag}$). When comparing multiple \hst\@ images at various epochs, these faint sources can be cross-matched with each other to recover their PMs and parallaxes to much fainter magnitudes. While the pipeline currently offers this feature, it is currently untested and optimized, but preliminary results suggest this approach will be fruitful. It may prove particularly useful in very sparse fields where the number of \gaia\@ sources in each \hst\@ image is small, but the shared source list between the \hst\@ frames is large. 

\subsection{Caveats} \label{ssec:pipeline_caveats}

One key assumption that is built in to \gaiahub\@ -- and therefore \bpm\@ -- comes from how \hst\@ sources are cross-matched with \gaia\@. An \hst\@ source is matched with a \gaia\@ source if they are nearest neighbours within some angular distance of each other (e.g. \gaiahub\@ default of 5 \hst\@ pixels, $\sim250~\mathrm{mas}$, though this can be changed), which ultimately sets upper limits on the sizes of PMs that can be measured. This works quite well in medium to low density regions where the stars are likely far enough from each other with respect to their motion between images. In high density regions or for fast moving stars, the closest pair of sources in successive images are not necessarily the correct matching. While future versions of our pipeline may adjust the cross-matching technique to reduce this confusion (e.g. rewind the \gaia\@ sources using \gaia-measured PMs, where they exist, before cross-matching and defining matches probabilistically), we stress that our work focuses on the low to medium density regions where the cross-matching assumptions are valid. 

As a back-of-the-envelope test, the cross-matching technique is a good assumption when the average distance between stars in an HST image is twice as large as the average position change from PM:
\begin{equation} \label{eq:good_crossmatch}
    \rho_*^{-1/2} > 2\cdot \Delta t \cdot \bar \mu.
\end{equation}
where $\rho_*$ is the stellar number density in area on the sky. We can use this equation to determine different limits for choices of time baselines, densities, and average PM sizes. For example, with ACS/WFC's $4096\times4096$ pixel detector with pixel scale of $50~\masppixel$, a time baseline of 15 years, and an average PM of $100~\maspyr$, we find that there would need to be greater than $4660$ stars in the image for there to be a significant amount of confusion in the cross-matching; as these numbers are similar to HST images in the COSMOS field, this implies that sparse regions in the halo have very low risk of incorrect cross-matching. In denser regions, like nearby galaxies, a time-baseline of 15 years and average PM of $5~\maspyr$ implies that a limiting stellar number density of $1/9~\mathrm{pixel}^{-2}=4.4\times10^{-5}~\mathrm{mas}^{-2}$. In practice, we would like the factor of 2 to be even larger (e.g. 5 or 10) to be safe, but this sets the threshold. Future versions of the pipeline will likely explore improvements to the cross-matching between \hst\@ and \gaia\@ using the posterior positions and motion measurements from \bpm.

\section{Validation with Synthetic Data} \label{sec:validation}

\begin{figure*}[ht]
\begin{center}
\includegraphics[width=\linewidth,height=6in,keepaspectratio, trim={0cm 0cm 0cm 0cm},clip]{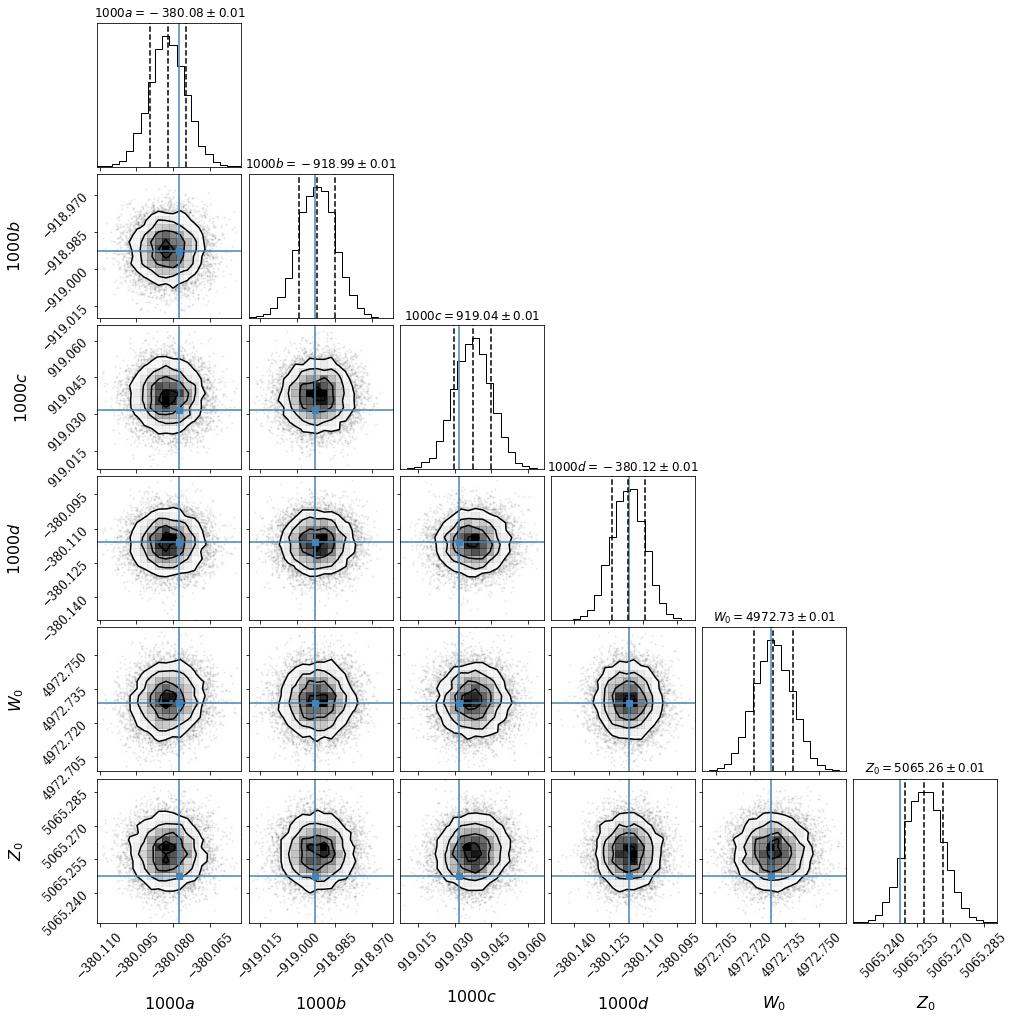}
\caption{Corner plot of the posterior samples (black points, lines, and histograms) of the transformation parameters when fitting an image with 200 stars with a 15 year time-baseline. The blue lines show the locations of the input parameters used to create the synthetic data, and the values of the transformation matrix $(a,b,c,d$) have been multiplied by 1000 for clarity. The chosen input parameters are representative of real transformation solutions measured by \gaiahub\@ for \hst\@ ACS/WFC images.}
\label{fig:fake_transform_distribution}
\end{center}
\end{figure*}

To understand how the pipeline performs in ideal conditions where we know the input transformation parameters and stellar motions perfectly, we first generate synthetic, COSMOS-like data. This process is described in detail in Appendix \ref{sec:generating_synthetic_data}. In summary, the synthetic sources are a mix of foreground thick disk stars and halo stars covering the $16<G<21.5~\mathrm{mag}$ range with realistic \gaia\@ measurements and uncertainties of position, parallax, and PM. Like with real \gaia\@ data, sources with $G>21~\mathrm{mag}$ have no \gaia-measured PMs or parallaxes. We create synthetic \hst\@ observations of these sources -- as well as the corresponding \gaiahub-like output catalogs that \bpm\@ expects to use for initial guesses of the transformation parameters -- while varying the numbers of sources per \hst\@ image and time baselines.
Finally, the data from the synthetic catalogs are analysed by \bpm. We emphasize that creating synthetic data with different configurations (e.g. transformation parameters, time baselines, stellar velocities and distance distributions) is quite straightforward using our technique, and it is not necessarily restricted to only \hst-like observations. Our method is a useful avenue for estimating the impact that future observations or telescopes can have on stellar motion measurements as well as designing best practices. 

\subsection{Recovering Transformation Parameters} \label{ssec:recover_transformation}

While the comparison with real data will follow in Section~\ref{sec:applications}, we begin by testing the pipeline on synthetic data to see how well we can recover the input transformation parameters and stellar motion (i.e. position, parallax, and PM) used to generate the synthetic \hst\@ image. Figure~\ref{fig:fake_transform_distribution} shows the posterior distributions on the transformation parameters that \bpm\@ measures for one synthetic \hst\@ image that has 200 sources and a 15~year time baseline from \gaia\@; the black points and histograms show the posterior draws, while the blue lines show the true values used to generate the image. The posterior distribution and the input values agree very strongly with each other. The chosen input transformation parameters are representative of real transformation solutions as measured by \gaiahub\@. 

\begin{figure}[ht]
\begin{center}
\includegraphics[width=\linewidth]{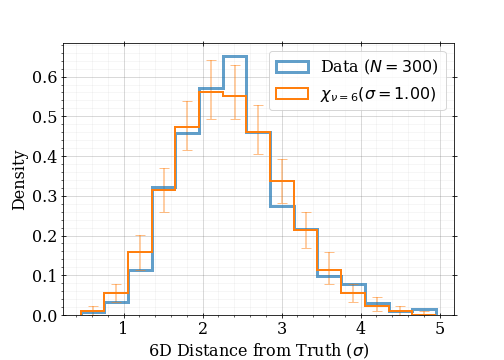}
\caption{6D distance of the transformation parameters (i.e. the parameters in Figure~\ref{fig:fake_transform_distribution}) from their true values for 300 realizations of synthetic \hst\@ images with time baselines from 5 to 15 years and number of sources from 5 to 200. The blue histogram shows the data, while the orange curve is the expected distribution; the agreement between these histograms show that the \bpm\@ pipeline does a good job of recovering the input transformation parameters within their uncertainties.}
\label{fig:fake_transform_dist_from_true}
\end{center}
\end{figure}

We repeat this process of generating synthetic images while changing the number of sources and the time baseline, and then measure how far the posterior distribution is from the truth. Specifically, we use the posterior samples of the transformation parameters (e.g. the black points in Figure~\ref{fig:fake_transform_distribution}) to define a 6D posterior mean vector, $\vec v$, and corresponding $6\times6$ covariance matrix, $\pmb V_{\vec v}$. The 6D distance between the posterior mean vector and the truth vector, $\vec v'$, is then defined using the Mahalanobis distance metric
\begin{equation} \label{eq:residuaL_distance}
    \begin{split}
        D = \left( (\vec v -\vec v')^T \cdot \pmb V_{\vec v}^{-1} \cdot (\vec v -\vec v')\right)^{1/2}.
    \end{split}
\end{equation}

\begin{figure*}[t]
\begin{center}
\includegraphics[width=\linewidth,height=4in,keepaspectratio, trim={0cm 0cm 0cm 3.5cm},clip]{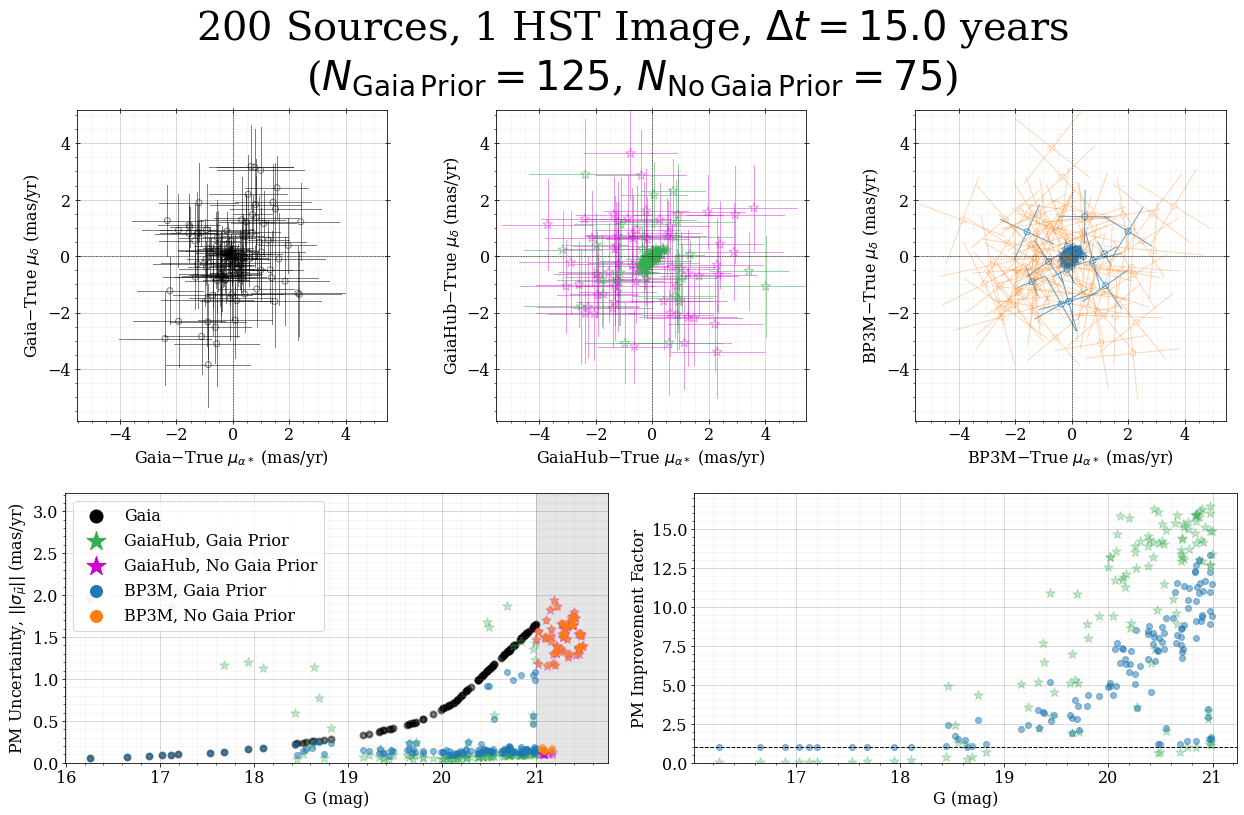}
\caption{Comparison of \gaia\@, \gaiahub\@, and \bpm\@ PMs for 200 synthetic COSMOS-like stars with a 15 year time baseline from \gaia\@. From left to right, The upper panels shows the difference of the \gaia\@, \gaiahub\@, and \bpm\@ PMs from the true PM values, and all axes share the same limits. In the \gaiahub\@ and \bpm\@ panels, points in green or blue show PMs of sources that are bright enough to have synthetic \gaia\@ parallax and PM measurements and magenta or orange points show PMs of sources that are too faint for \gaia\@ to have measurements. The bottom two panels compare the PM vector uncertainty size between \gaia\@, \gaiahub\@, and \bpm\@, with the right panel showing the division of the black points by the green and blue points from the left panel.}
\label{fig:fake_pm_comparison}
\end{center}
\end{figure*}

This is analogous to the 1D case of dividing the absolute difference between a mean value and the truth by the uncertainty. Likewise, the units of $D$ can be thought of as the number of $\sigma$ between the truth and mean, and we will use this definition (for varying dimensions of $\vec v$) when studying residuals throughout this work. When we analyse 300 synthetic \hst\@ images with time baselines between 5 and 15~years from \gaia\@ and 5 to 200 sources per image, we get the blue histogram in Figure~\ref{fig:fake_transform_dist_from_true}; the orange distribution shows the expected outcome -- namely, a $\chi$ distribution with 6 degrees of freedom and a scale of 1.0 -- which agrees well with the measured outputs. This result shows that the pipeline does a good job of recovering the input transformation parameters within the posterior uncertainties.

\subsection{Recovering Motions} \label{ssec:recover_PMs}

\begin{figure}[t]
\begin{center}
\includegraphics[width=\linewidth]{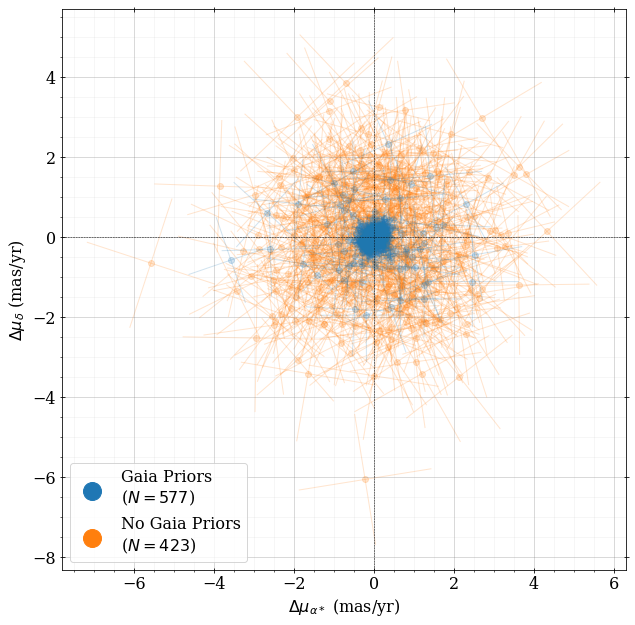}
\caption{Comparison of \bpm-measured posterior PM's with the truth for 5 sets of synthetic \hst\@ images, each with 200 sources and a 15 year time baseline from \gaia\@. The points are colored by whether the \bpm\@ sources have \gaia\@ parallax and PM priors (blue), or are too faint to have \gaia\@ parallax and PM priors (orange). The uncertainty lines are the eigenvalues of the posterior covariance matrix, scaled so that each line corresponds to 68\% probability. All the points are clustered around (0,0), implying that \bpm\@ recovers trustworthy PMs.}
\label{fig:fake_VPD}
\end{center}
\end{figure}

\begin{figure}[t]
\begin{center}
\includegraphics[width=\linewidth]{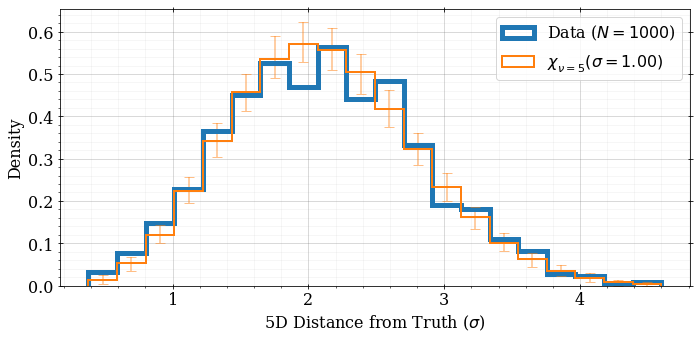}
\caption{Comparison of uncertainty-scaled distance of the 5D posterior vectors (i.e. 2D position, parallax, and 2D PM) from the truth. The blue histogram represents the \bpm\@ data for the 1000 synthetic stars in Figure~\ref{fig:fake_VPD}, and the orange histogram is the expected distribution ($\chi$ with 5 degrees of freedom and a scale of 1.0). The agreement between these two curves is evidence that the pipeline is recovering good posterior 5D vectors with realistic uncertainties.}
\label{fig:fake_5D_dists}
\end{center}
\end{figure}

\begin{figure*}[t]
\begin{center}
\includegraphics[width=\linewidth,height=4in,keepaspectratio, trim={0cm 0cm 0cm 3.5cm},clip]{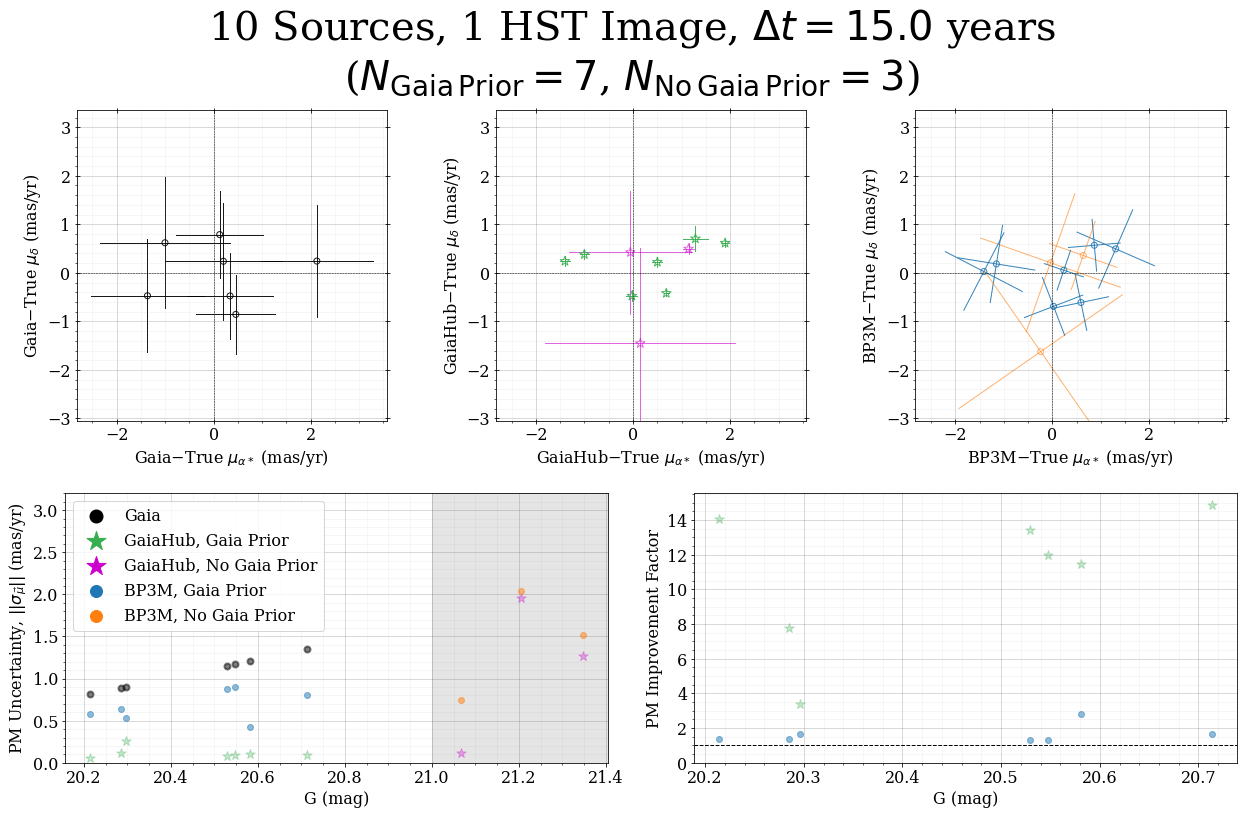}
\caption{Same as Figure~\ref{fig:fake_pm_comparison}, but with 10 synthetic COSMOS-like stars. The top panels share axis limits. The \bpm\@ PM improvement factor is less significant than in the bottom right panel of Figure~\ref{fig:fake_pm_comparison}, owing to the larger uncertainty in the transformation parameter fitting because of the smaller number of sources.}
\label{fig:fake_pm_comparison_n010}
\end{center}
\end{figure*}

Next, we analyse how well the pipeline is able to recover the true motions of the sources in each synthetic image of MW halo stars. We also analyse the synthetic data using a \gaiahub-like approach, though we cannot pass the exact synthetic data to \gaiahub\@ because it requires raw \hst\@ images that are not currently included in the simulations. Still, the \gaiahub\@ analysis of synthetic data is similar to what \gaiahub\@ would produce after iteration on the transformation parameter fitting (e.g. using the \texttt{--rewind\_stars} option), which makes it the same as the analysis mode used to analyze the real COSMOS data in Figure~\ref{fig:COSMOS_gaiahub_comp}.

An example comparison of the PMs for \gaia\@, \gaiahub\@, and \bpm\@ is shown in Figure~\ref{fig:fake_pm_comparison} for the same synthetic image that is considered in Figure~\ref{fig:fake_transform_distribution}, where in each panel we compare the observed PMs to the known truth. When showing \gaiahub\@ and \bpm\@ results (top middle and right panels), we color the sources by whether there are \gaia\@-measured parallaxes and PMs to be used as priors; green or blue points designate sources with (``Gaia Prior'') and magenta or orange points designate sources without (``No Gaia Prior''). To be precise, all of the sources have \gaia-measured priors on their position at J2016.0, but not all sources have \gaia\@ priors on their parallaxes and PMs. 

The first key takeaway from this figure is that the \gaiahub\@ and \bpm\@ PMs are clustered closer to the true values (i.e. the origin in the upper panels) when compared to \gaia\@ alone: the black points in the \gaia\@ panel represent the same sources as the green points in the \gaiahub\@ panel and the blue points in the \bpm\@ panel. In the \gaiahub\@ panel, the green points near the origin have a systematic trend along the $y=x$ line that is not apparent in the \bpm\@ panel. Investigating this trend using different synthetic data with different PM populations, it seems that large PMs are likely the cause. When focusing on smaller-magnitude PM populations (e.g. the extragalactic populations that \gaiahub\@ was developed for), this systematic trend disappears. As a result, the PM uncertainties in the \gaiahub\@ panel (and the corresponding points in the lower panels) need to be $\sim2$ times larger on average to explain the difference between the measured PMs and the truth, whereas the \bpm\@ PM uncertainties completely capture any differences with the truth. While \gaiahub\@ and \bpm\@ both show improvements over \gaia\@ alone, \bpm\@ handling of large PMs (e.g. as expected in the MW halo) and the accuracy of the PM uncertainties help motivate why \bpm\@'s creation was necessary. 

Next, the bottom panels show that the size of the PM uncertainty has decreased significantly when \hst\@ information is combined with \gaia\@ using \bpm\@; the left panel compares the PM uncertainties, while the right panel shows how much smaller the \bpm\@ PM uncertainties are, compared to the \gaia\@ PM uncertainties, as a function of magnitude. As mentioned in the previous paragraph, the \gaiahub\@ uncertainties would need to be a factor of 2 larger than the values shown in this figure to explain the offsets of the measured PMs from the truth. Overall, this figure shows that combining the \hst\@ and \gaia\@ data with \bpm\@ not only increases PM precision, it also increases accuracy.
To be clear, \bpm\@ is not simply shrinking the \gaia\@ PM distribution around the same mean as a result of an increased time baseline and number of images, it is truly improving the astrometric solution of each source. 

The particular pattern of PM uncertainties (i.e. a lower branch around $0.2~\maspyr$ and an upper branch around $1.5~\maspyr$) is a consequence of our choices in modelling the synthetic data, which is detailed in Appendix~\ref{sec:generating_synthetic_data}. As a brief summary, the \hst\@ position uncertainties are based on real \gaiahub\@-analysed sources as a function of magnitude in COSMOS. At a given magnitude, some of the sources have well-measured positions, while others are less constrained, which leads to the PM uncertainty bifurcation shown in Figure~\ref{fig:fake_pm_comparison}.

When considering the uncertainty on a vector, there are a few different approaches one can take. In this work, we are concerned with 2D vectors (e.g. positions, PMs), 3D vectors (e.g. parallax plus 2D PM), and 5D vectors (i.e. 2D position, parallax, and 2D PM) of the different motion components. Because the different components of these vectors can have differing units, and because there can be substantial correlations in the covariance matrices we measure, we choose not to use the standard metric of the quadrature sum of the individual uncertainties of each component. Instead, when we compare vectors that have associated covariance matrices, we are interested in how much the size of that covariance matrix has changed, which includes the effect of correlation. We define the size of a vector's uncertainty to be the determinant of the covariance matrix to the $1/(2d)$ power, where $d$ is the vector's number of dimensions:
\begin{equation}
    \label{eq:vector_uncertainty}
    || \sigma_{\vec v} || = |\pmb V_{\vec v} |^{1/(2d)}
\end{equation}
where $\pmb V_{\vec v}$ is the covariance matrix of $\vec v$. With this definition, the area/volume created by the vector's covariance matrix, i.e. the determinant $|\pmb V_{\vec v}|$, is equal to the area/volume of a purely diagonal covariance matrix:
\begin{equation*}
    \pmb V_{|| \sigma_{\vec v} ||} = \left(\begin{matrix}
    || \sigma_{\vec v} ||^2 & 0 & \dots & 0\\
    0 & || \sigma_{\vec v} ||^2 &  & \vdots\\
    \vdots & & \ddots & 0\\
    0 & \dots & 0 & || \sigma_{\vec v} ||^2\\
    \end{matrix}\right).
\end{equation*}
In this case, the resulting uncertainty size allows for the correlations between parameters to impact the certainty about the vector's position, which serve to shrink the volume defined by that covariance matrix. For the remainder of this work, where we compare the uncertainty sizes of different vectors between \gaia\@ and \bpm\@, we will be using the vector uncertainty size definition of Equation~\ref{eq:vector_uncertainty}. 

As an illustrative example, a highly-correlated measurement might have large uncertainties in all of the individual vector components, but the probability distribution implied by its covariance matrix covers only a small volume of parameter space owing to the high correlation between the components. Here, the quadrature sum of individual uncertainties would yield a large result, implying we know little about the true value of the vector, whereas the vector uncertainty size in Equation~\ref{eq:vector_uncertainty} would return a small value. This small value tells us that the volume of possible values that the vector can occupy is quite small because of the relationship between the dimensions. For a 2D vector $\vec v^T = (x,y)$, the vector uncertainty size is given by:
\begin{equation}
\langle \sigma_{\vec v} \rangle = \left(\sigma_x^2 \cdot \sigma_y^2 \cdot \left[ 1-\rho_{x,y}^2\right]\right)^{1/4}
\end{equation}
where $\sigma_x$ and $\sigma_y$ are the corresponding uncertainty in each of the vector components and $\rho_{x,y}$ is the correlation coefficient between $x$ and $y$. 

To account for differences that individual realizations might have on the posterior \bpm\@ PMs, we repeat the measurements of Figure~\ref{fig:fake_pm_comparison}; that is, we create 5 separate synthetic \hst\@ images that have 200 sources and a time baseline of 15~years. For the different realizations, the list of randomly chosen sources all come from the same synthetic catalog of COSMOS-like data, but each realization will have a slightly different set of true PMs, parallaxes, positions, magnitudes, and numbers of sources with/without \gaia\@ priors. The differences of the posterior PMs from the truth for these $5\times200=1000$ sources are shown in Figure~\ref{fig:fake_VPD}, where the sources with \gaia\@-measured parallaxes and PMs are in blue and the converse are in orange. As expected for a well-behaved pipeline, the difference distribution clusters around the origin, and we see that the sources with \gaia\@ priors on their motion have smaller posterior uncertainties than those without.

Of course, the PMs are only 2 of the 5 dimensions of the vector measured for each source, so we also compare the 5D vector (2D position, parallax, 2D PM) to the truth, again using the distance definition of Equation~\ref{eq:residuaL_distance}; this distribution for the 1000 stars in Figure~\ref{fig:fake_VPD} is given as the blue histogram in Figure~\ref{fig:fake_5D_dists}, and it shows remarkable agreement with the expected $\chi$ distribution with 5 degrees of freedom and a scale of 1. These figures show that the pipeline recovers the true positions, parallaxes, and PMs of all sources within their posterior uncertainties. 

Finally, we explore the impact that the number of sources in an image have on the PM improvement factor when comparing \bpm\@ to \gaia\@. An example analysis of a synthetic \hst\@ image with 10 sources and a 15~year time baseline is given in Figure~\ref{fig:fake_pm_comparison_n010}. While the \bpm\@ pipeline again finds good agreement with the truth, the median PM improvement factor over \gaia\@ alone is only 1.39 for the sources in this synthetic image, which is much smaller than the median factor of 8.3 seen for the sources in the $20.5<G<21~\mathrm{mag}$ range of Figure~\ref{fig:fake_pm_comparison}. This is largely because the smaller number of sources aren't able to place as strong a constraint on the transformation parameters' distribution, and the larger transformation parameter distribution propagates to the PM distributions of individual sources. 

Comparing the \gaiahub\@ and \bpm\@ panels, there is good agreement between the PM means for this particular realization of synthetic data, which should not be taken for granted considering the large \gaiahub\@ outliers that can appear in real COSMOS data (e.g. $\sim10\%$ of the PMs in Figure~\ref{fig:COSMOS_gaiahub_comp}). When we compare the PM uncertainties, however, we see significant differences between the two pipelines. In fact, the \gaiahub\@ uncertainties would need to be increased by a factor of 8 to explain the differences with the true PMs, while the \bpm\@ PMs agree with the truth within their uncertainties. Some of the PM uncertainty differences come from the fact that the \gaiahub\@ analysis is only able to propagate realistic transformation solution uncertainties to the final PMs if there are several \hst\@ images of the same field, which is not the case when considering a single synthetic image. The impact of the transformation uncertainty is likely negligible for relatively large numbers of sources \citep[e.g. $N>100$, as discussed in][]{delPino_2022}, but it becomes quite important in sparse fields and when there are only a handful of repeat \hst\@ observations of the same field (e.g. $N=4$ for many fields in COSMOS).

While it is fairly straightforward to estimate how much the \gaiahub\@ uncertainties need to be inflated by using comparisons with the \gaia-measured PMs where they exist, it is much more difficult to assess and correct for the presence of systematics (e.g. as seen in the \gaiahub\@ PMs of Figure~\ref{fig:fake_pm_comparison}) in sparse fields. Fortunately, these sparse halo conditions are what \bpm\@ was developed for, and the synthetic data analysis provide evidence that the pipeline provides accurate PMs and uncertainties here. 

We leave out example figures exploring the effect that a changing time baseline has on the posterior PMs because it follows expectations exactly; namely, for the same position uncertainty between two images, a smaller time baselines leads to larger PM uncertainties \citep[e.g. Equation 2 of][]{delPino_2022}.

\subsection{Optimizing Observing Strategies for Positions and Parallaxes} \label{ssec:recover_parallaxes}

Our synthetic analysis also allows us to measure the improvement factors in the parallaxes and 2D position vectors of the sources. These results show, not unexpectedly, that the precision of these measurements is strongly tied (but not limited) to the magnitude of the source, the time baseline between images, the number of sources per image, and the effect of the position on the sky on the shape and orientation of the parallax motion. While we cannot change the time baselines of archival \hst\@ images, we can explore how changes in the observations times would have impacted the resulting \bpm\@ parallaxes and positions. These lessons are particularly useful for planning of future observations using any telescope. 
\begin{figure*}[h]
\begin{center}
\includegraphics[width=0.9\linewidth]{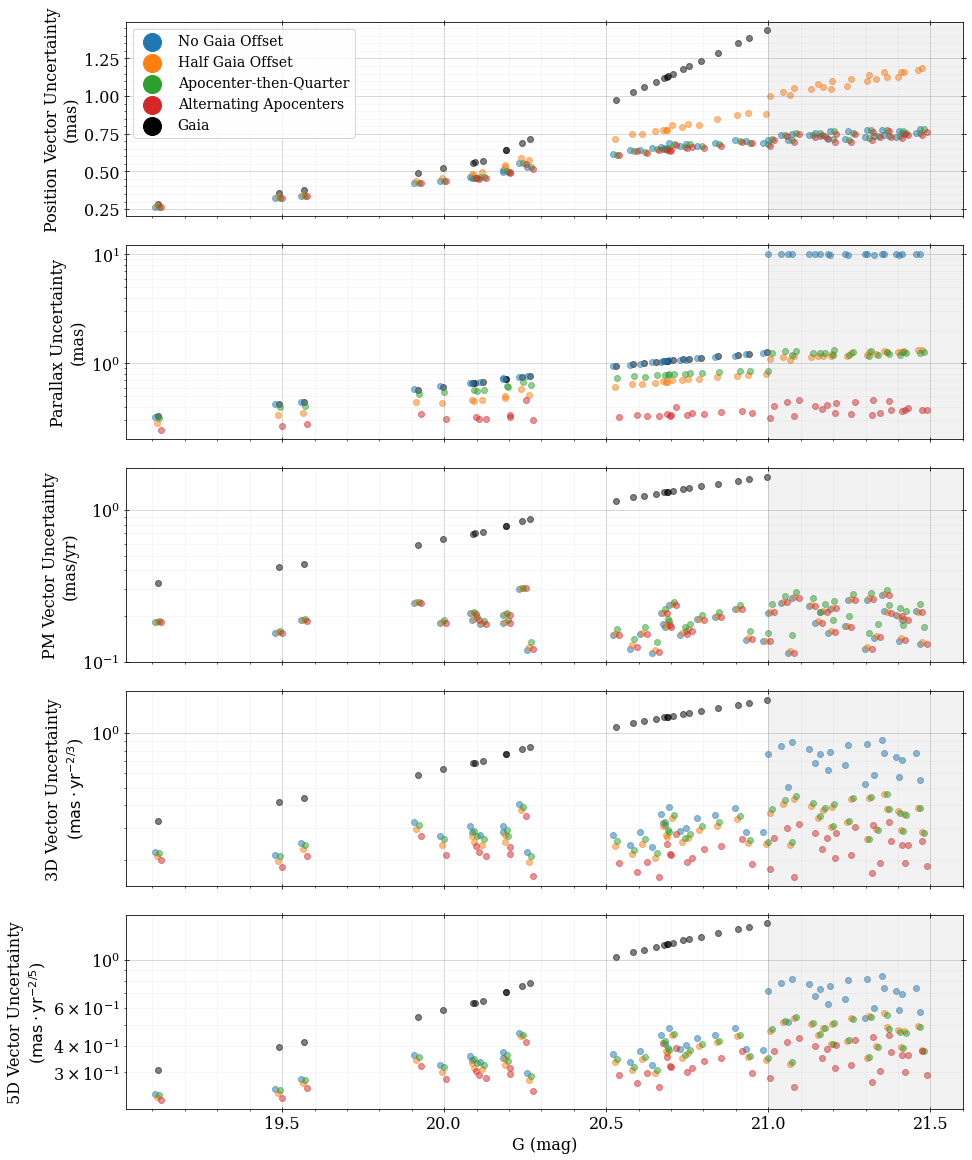}
\caption{Comparison of the uncertainties on position, parallax, and PM for 50 synthetic \hst\@-observed stars in the COSMOS field. The points are colored by different observation strategies (i.e. time baseline offsets from \gaia\@) with the legend names described in the text. Note that some of the y-axes are in log scale while others are linear. The two bottom-most panels correspond to the 3D vector of parallax and PM, and the 5D vector of position, parallax, and PM; the units of these axes are a result of our definition of vector uncertainty in Equation~\ref{eq:vector_uncertainty}.}
\label{fig:synthetic_HST_uncertainties}
\end{center}
\end{figure*}

As before, we create synthetic \hst\@ images of COSMOS stars, but in this analysis, we do not let the \hst\@ position uncertainty of sources change as a function of magnitude based on \gaiahub\@ examples (e.g. as discussed in Appendix~\ref{sec:generating_synthetic_data}). Instead, we choose to set the \hst\@ position uncertainty to 0.01~pixels, which is the centroid uncertainty that previous studies have measured for well-behaved stars \citep[e.g.][]{Bellini_2011}. We adopt the single, well-behaved value to limit the effect that any particular choice/realization of position uncertainties may have on the conclusions we make about optimal observing strategies. Similarly, we choose to use 50 stars in each synthetic \hst\@ image to reduce the impact of transformation solution uncertainty as seen in low-source-count images.

Our initial tests suggest that there is minimal improvement on the \gaia-measured parallax or position vectors when using only one \hst\@ image (i.e. one \hst\@ in combination with \gaia\@). Additionally, multiple \hst\@ images taken very nearby in time (e.g. the multiple exposures taken to guard against cosmic rays) effectively act as a single epoch and are useful for improving the position uncertainty, but don't offer much information to the parallax. When multiple \hst\@ images with common targets and substantial enough time baselines are analysed together, then the parallaxes and positions begin to show improvement. To better understand how parallax and position precision depend on observation times, we choose to simulate 3 epochs of COSMOS observations with \hst\@, with time offsets from \gaia\@ near 12, 8, and 4 years. Specifically, we make the observations occur at the following times:
\begin{enumerate}
    \item \textbf{No Gaia Offset}: All \hst\@ observations at year-multiples of the \gaia\@ observation time, \newline $\Delta t=\left[-12.0,-8.0,-4.0\right]~\mathrm{years}$;
    \item \textbf{Half Gaia Offset}: All \hst\@ observations at half-a-year offset from the \gaia\@ date, $\Delta t=\left[-11.5,-7.5,-3.5\right]~\mathrm{years}$;
    \item \textbf{Apocenter-then-Quarter}: The earliest \hst\@ observations at the parallax orbit apocenter farthest from \gaia\@, then offset by a quarter year for the other epochs, \newline $\Delta t=\left[-11.61,-7.36,-3.86\right]~\mathrm{years}$;
    \item \textbf{Alternating Apocenters}: The earliest \hst\@ observations at the parallax orbit apocenter farthest from \gaia\@, then alternating offsets by half a year for the other epochs, \newline $\Delta t=\left[-11.61,-7.11,-3.61\right]~\mathrm{years}$.
\end{enumerate}

The resulting position, parallax, PM uncertainties from analysing these \hst\@ observations concurrently is given in Figure~\ref{fig:synthetic_HST_uncertainties}, where the two bottom-most panels show the 3D (i.e. parallax, PM) and 5D (i.e. position, parallax, PM) vector uncertainties. This figure reveals many important lessons about the impact that relatively small time offsets of observations within a year have on the stellar measurements. 

From the top three panels, we see that different observing strategies lead to significant differences in the resulting position and parallax precisions, though there are minimal differences between the PM uncertainties. This implies that improvements in parallax and position can be achieved with no cost to the PM precision. We emphasize, however, that this is only after analysing all 3 \hst\@ epochs together; when considering a single epoch at a time, sources without \gaia\@ priors (i.e. $G>21~\mathrm{mag}$) do indeed have significantly larger PM uncertainties if there is an offset from the \gaia\@ date as a result of the degeneracy between PM and parallax motions. Our results suggest that completed surveys with multiple \hst\@ epochs may be able to improve position and parallax measurements for free through careful design. 

Of the observing strategies we consider (including many that are not represented in Figure~\ref{fig:synthetic_HST_uncertainties}), the best choice for improving positions and parallaxes was to use an alternating-parallax-orbit-apocenters approach, which is the 4th option in the list above and the red points in the figure. This strategy suggests that the parallax is most improved -- and therefore the degeneracy between PM and parallax motions most disentangled -- when we consider the average parallax orbit of the sources in a given field (e.g. the blue orbit in Figure~\ref{fig:parallax_motion} for COSMOS). The parallax ellipse can be used to identify the points in time that are most distant from each other (i.e. the extremes of the semi-major axis), which are therefore the easiest to tell apart for a given position uncertainty. Then, the best approach is to first observe at the time that is most offset from the \gaia\@ observation time, and subsequent observations alternate by half a year such that they jump back and forth between the two apocenters. This approach also appears to have a large improvement on the position uncertainty, likely caused by the repeated sampling of the source's position at the same time in the parallax orbit. 

\section{Applications with Real Data} \label{sec:applications}

We test the pipeline using real data in four key ways: (1) using well-studied nearby dwarf spheroidal (dSph) galaxies, (2) using cross-matches with an external QSO catalog, (3) using cross-matches with a PM catalog in COSMOS derived from multi-epoch \hst\@ imaging, and (4) using the full $\sim2000$ unique COSMOS sources presented in Section~\ref{ssec:COSMOS_testbed}.

\subsection{Comparison with Nearby Dwarf Spheroidals} \label{ssec:dSph_comp}

\begin{figure*}[h]
\begin{center}
\includegraphics[width=0.9\linewidth,height=3.5in,keepaspectratio, trim={0cm 0cm 0cm 3.5cm},clip]{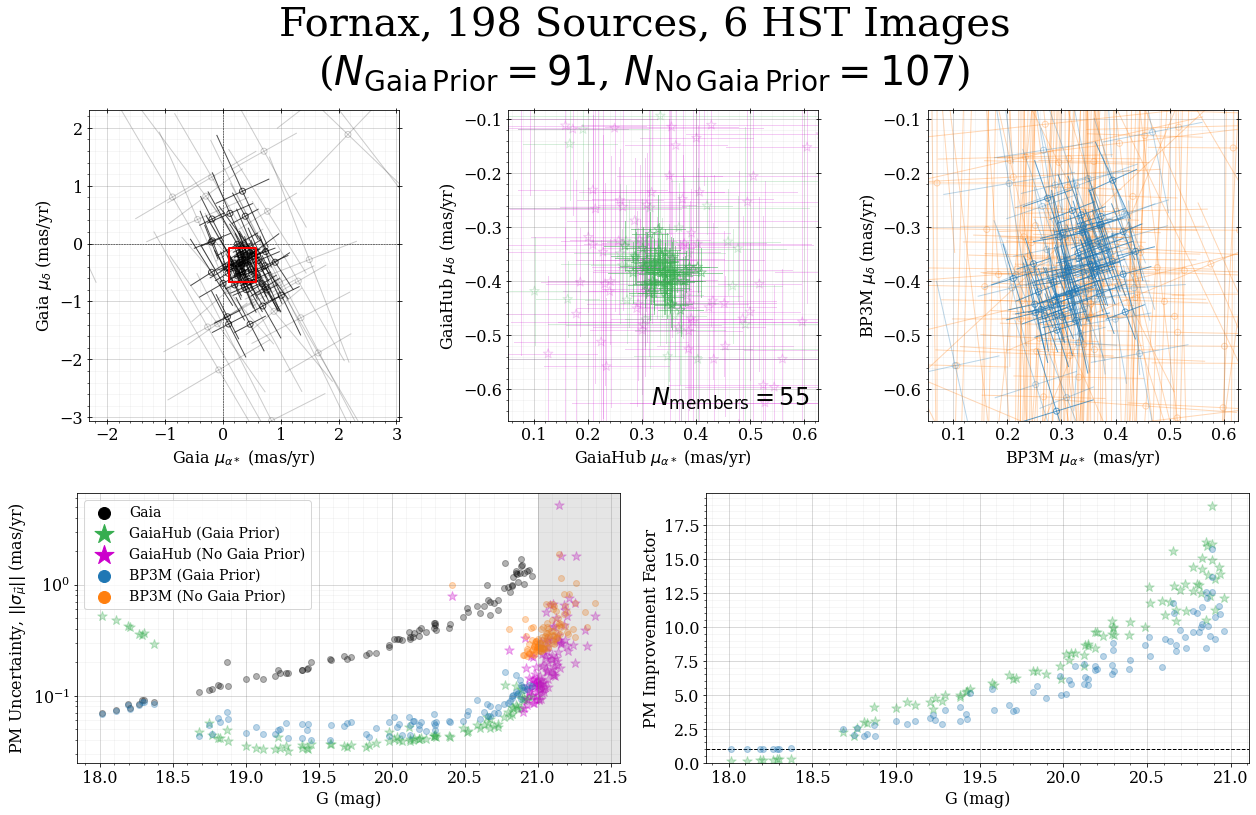}
\caption{Comparison of PMs using 6 \hst\@ images of Fornax dSph analysed concurrently; these \hst\@ images were all taken using ACS/WFC in the F775W filter at the same epoch and same RA and Dec. The exposure time of each image is $\sim250~\mathrm{sec}$ and the time offset from \gaia\@ is 12.8~years. The panels are similar to Figure~\ref{fig:fake_pm_comparison}, except the upper panels show measured PM values instead of offsets from known PMs of synthetic data; the upper left panel shows \gaia\@ measurements, the middle panel shows \gaiahub\@-measured PMs, and the right panel shows \bpm\@ posteriors. In all of the top panels, the 55 sources identified as cluster members by \citet{delPino_2022} are plotted with more-opaque points. Sources in the \gaiahub\@ and \bpm\@ panels are colored by whether \gaia-measured PMs exist (i.e. 91 sources with \gaia\@ PMs and 107 without \gaia\@ PMs). The \gaiahub\@ and \bpm\@ panels cover the same range of PM values to visually compare the PM clustering, with that range shown with a red box in the \gaia\@ panel. \label{fig:fornax_pm_comparison}}
\includegraphics[width=0.9\linewidth,height=3.5in,keepaspectratio, trim={0cm 0cm 0cm 3.5cm},clip]{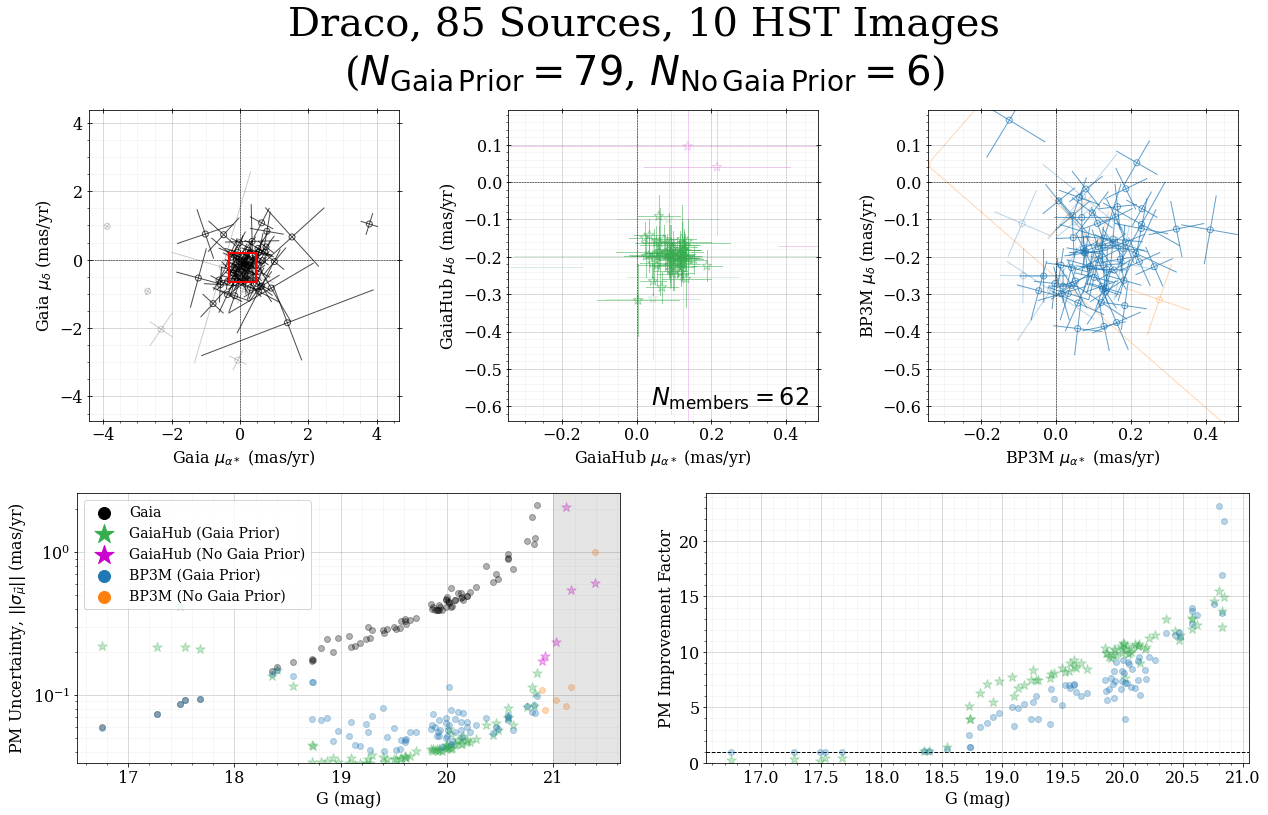}
\caption{Same as Figure~\ref{fig:fornax_pm_comparison}, but analyzing 10 \hst\@ images of Draco concurrently. These \hst\@ images were all taken using ACS/WFC in the F606W filter at the same RA and Dec, with an average exposure time of $\sim215~\mathrm{sec}$. The images span three epochs, with time offsets from \gaia\@ of 11.2, 9.2, and 2.2~years. There are 79 sources with \gaia\@ PMs and 6 without \gaia\@ PMs. \citet{delPino_2022} identifies 62 member stars in this sample. \label{fig:draco_pm_comparison}}
\end{center}
\end{figure*}

\begin{figure}[ht]
\begin{center}
\includegraphics[width=\linewidth]{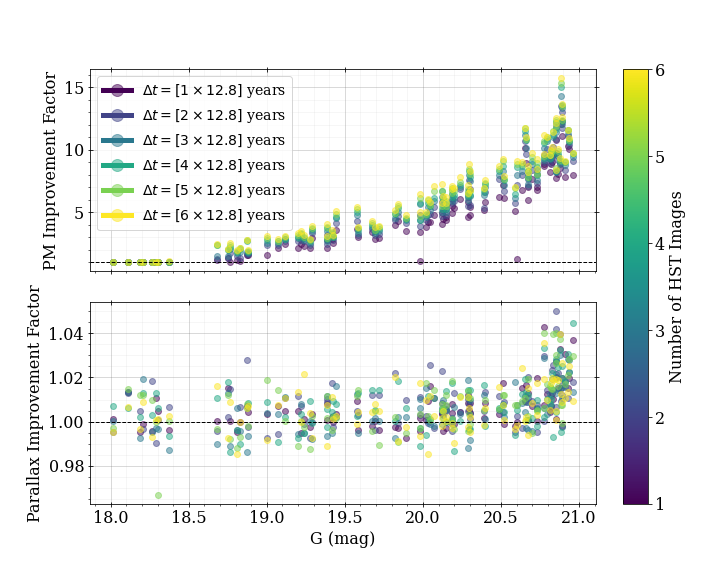}
\caption{Comparison of parallax and PM improvement factor to \gaia\@ (i.e. size of \gaia\@ parallax or PM uncertainty divided by size of \bpm\@ parallax or PM uncertainty) for Fornax dSph as a function of the number of \hst\@ images used. These measurements concern the same data presented in Figure~\ref{fig:fornax_pm_comparison}. While the most significant improvement happens when combining a single \hst\@ image with \gaia\@ (i.e. median factor of $\sim8.6$ for $20.5<G<21~\mathrm{mag}$), there is still a $\sim14\%$ improvement for the faintest sources between considering one \hst\@ image and considering six. There is also a very slight improvement on the parallax uncertainties at the faintest magnitudes (for one image, median factor of $\sim1.01$ for $20.5<G<21~\mathrm{mag}$), though the \hst\@ observations occurring at the same approximate time make it difficult to improve the parallax precision.}
\label{fig:fornax_pm_improvement_per_image}
\end{center}
\end{figure}

\begin{figure}[ht]
\begin{center}
\includegraphics[width=\linewidth]{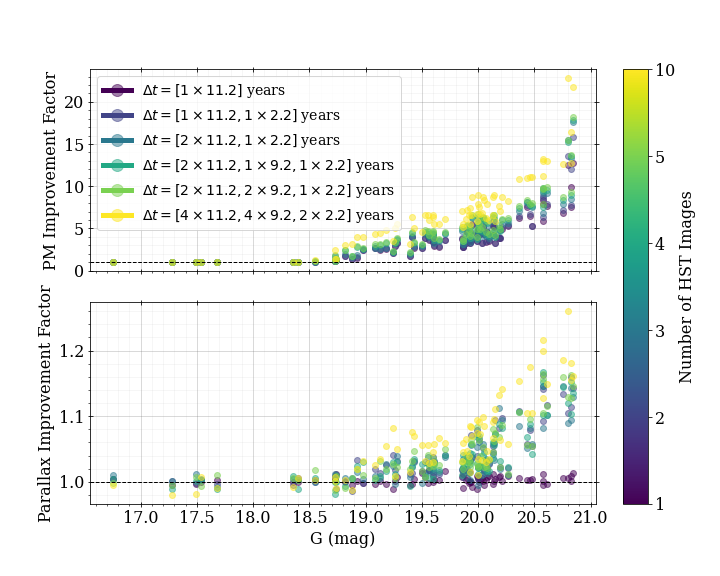}
\caption{Same as Figure~\ref{fig:fornax_pm_improvement_per_image}, but for Draco instead of Fornax. These measurements concern the same data presented in Figure~\ref{fig:draco_pm_comparison}. Note that the number of images jumps from 5 to 10 on the colorbar. In contrast to the Fornax results, there is a more-noticeable improvement in the parallax uncertainties at the faintest magnitudes when the number of images increases (e.g. $\sim18\%$ median improvement in precision for $20.5<G<21~\mathrm{mag}$ when $N_{im}=10$). This is likely a result of the \hst\@ images occurring at different epochs.}
\label{fig:draco_pm_improvement_per_image}
\end{center}
\end{figure}

We analyse \hst\@ images using \bpm\@ in nearby dSph galaxies to see how the posterior PMs and parallaxes compare to \gaia\@. This serves as a test with real data, as well as more proof that \bpm\@ is improving the astrometric solutions for the sources it measures. We also show how analysing multiple \hst\@ images together impact the resulting PM and parallax precision. 

A comparison of \gaia\@, \gaiahub\@, and \bpm\@ measured PMs is presented in Figure~\ref{fig:fornax_pm_comparison} for the Fornax dwarf spheroidal. For the \bpm\@ PMs, 6 \hst\@ images at the same epoch (time baseline from \gaia\@ of 12.8~years) with a total of 198 unique sources are analysed together. The \gaiahub\@ PMs are the same as in presented in \citet{delPino_2022} from analysis of the same 6 \hst\@ images, but in this case, a co-moving sample has been defined before transformation fitting to identify likely cluster members. While this figure is similar to Figure~\ref{fig:fake_pm_comparison}, the top panels are slightly different in that they show PM measurements instead of offsets from known PMs of synthetic data. Sources without \gaia-measured parallaxes and PMs are colored orange in the \bpm\@ panel, which show larger uncertainties than the sources with \gaia\@ priors in blue. The axes of the \gaiahub\@ and \bpm\@ panels have been set to cover the same range of values for easier visual comparison, and this range is $\sim10$ times smaller than in the \gaia\@ panel. As before, the lower panels compare the PM uncertainties. Figure~\ref{fig:draco_pm_comparison} shows another PM comparison, this time using 10 \hst\@ images of the Draco dwarf spheroidal, taken at 3 different epochs (baselines of 11.2, 9.2, and 2.2~years from \gaia\@). 

First, these figures show a significant tightening of the PMs when \gaia\@ and \hst\@ are combined using either \gaiahub\@ or \bpm\@. This suggests that the increase in PM precision over \gaia\@ alone is not at the cost of PM accuracy. In comparing the PM uncertainties, we find that the \gaiahub\@ values are slightly smaller than those of \bpm\@, but this is likely because \bpm\@ propagates transformation parameter uncertainties to the final PMs and \gaiahub\@ uses a single set of values for the transformation solution. We also see that the \gaiahub\@ PMs form visually-tighter PM knots than the \bpm\@ PMs, though we would expect them to be more similar based on the size of their PM uncertainties. These differences suggest that there are systematics in the \bpm\@ PMs that are not present in the \gaiahub\@ PMs. In testing possible explanations, we find that the \bpm\@ results are able to look more like the \gaiahub\@ ones when we restrict the samples to only the cluster members from \citet{delPino_2022}. Theoretically, the \bpm\@ statistics should not allow the presence of non-member sources to impact the PMs of member stars -- in fact, their presence should help improve the transformation solution -- but we find that this is not the case. 

Some further testing suggests the likely culprit comes from \bpm\@'s estimate of the source position uncertainty in the \hst\@ images. As discussed previously, \bpm\@ uses the same \hst\@ position uncertainties as \gaiahub\@, and this is calculated as a fraction of a PSF quality of fit statistic. When analysing a kinematically-cleaned, co-moving population of stars (i.e. the case that \gaiahub\@ was designed for), this estimate of uncertainty performs well because the sample is restricted to faint sources that follow a linear relation between the quality of PSF fitting and \hst\@ position uncertainty. The validity of this approximation is evident from the resulting \gaiahub\@ PMs and the results when \bpm\@ uses the same subsample. However, this uncertainty approximation can occasionally underestimate the true uncertainty -- especially for bright-but-not-saturated sources -- leading to cases where the transformation solution becomes torqued away from the truth by a few unfairly-high-weighted sources. As more \hst\@ images are analysed together (e.g. as we present here with the dSph data), the amount that \hst\@ information contributes to the posterior measurements begins to outweigh the \gaia-measured priors, and it becomes more important for the \hst\@ uncertainties to be more accurate. It will be important for future versions of \bpm\@ to identify other methods of measuring the \hst\@ position uncertainties to be as statistically robust as possible. 

For all the \bpm\@ analyses presented in this work, we emphasize that we use all \gaia\@ sources found in each \hst\@ image to test the general nature of the pipeline. Our goal is to retain as many sources as possible for transformation fitting, which is especially important in sparse field cases. For \bpm\@, foreground bright stars -- which often have well-measured \gaia\@ positions, parallaxes, and PMs -- that do not belong to the distant populations of interest are often useful anchors for the transformation fitting. 

We next repeat the \bpm\@ analysis while varying the number of \hst\@ images considered together to explore the effect this has on the posterior PMs and parallaxes. The resulting improvement factors as a function of magnitude are given in Figures~\ref{fig:fornax_pm_improvement_per_image} and \ref{fig:draco_pm_improvement_per_image} for Fornax and Draco respectively, with the legend displaying the time baselines from \gaia\@ for the different sets of \hst\@ images. In the case of Fornax, the most significant improvement in PM occurs when analysing a single \hst\@ image with \gaia\@ (i.e. a median PM precision increase by a factor of $\sim8.6$ for $20.5<G<21~\mathrm{mag}$), though there is still an additional improvement when the remaining five \hst\@ images are brought in ($\sim14\%$ improvement for for $20.5<G<21~\mathrm{mag}$). The parallax uncertainties, however, do not improve very significantly for any of the images, though there is a slight, $\sim1\%$, precision increase at the faintest magnitudes. This is likely because the \hst\@ images are taken at a single epoch and the time baseline is quite close to a one year multiple of \gaia's observation date (i.e. only offset by 20\% of a year); this means that the parallax orbit is being sampled repeatedly in a small area at the same approximate time, which does not lead to significant increases in parallax precision. 

For Draco, increasing the number of \hst\@ images does substantially improve the PMs for each image added; considering 10 \hst\@ images together with \gaia\@ improves the PM precision by a median factor of $\sim13$ for $20.5<G<21~\mathrm{mag}$ compared to a median factor of $\sim7.8$ when only one \hst\@ image is used. The parallaxes also see a noticeable change in precision as a function of \hst\@ images analysed. While the time offsets from \gaia\@ are again only $\sim20\%$ of a year, the \hst\@ images are spread over three epochs. In the case of Fornax, we effectively have measurements of positions for each source at two different times (i.e. a single epoch of \hst\@ images plus \gaia\@) while the Draco data have measurements of positions at four different times; for the Draco stars then, there is smaller set of parallax and PM combinations that can explain those four position measurements per source than if there were only two position measurements. It is likely that the improvement in the parallax precision also contributes to the increase in PM precision because the degeneracy between the different types of apparent motion become less entangled. However, the time offsets of the \hst\@ images from \gaia\@ and from each other are not optimally spaced in the parallax orbit (i.e. near the apocenters), resulting in a median parallax precision increase by a factor of $\sim1.18$ for the faintest magnitudes ($G>20.5~\mathrm{mag}$).

\subsection{Comparison with QSOs} \label{ssec:QSO_comp}

\begin{figure}[ht]
\begin{center}
\includegraphics[width=\linewidth]{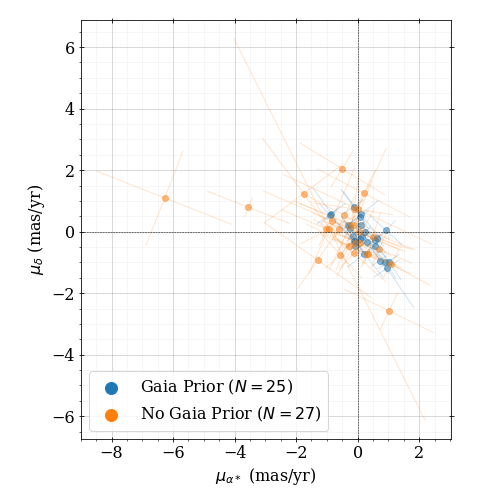}
\caption{PMs of the 52 sources in the COSMOS field that are nearest to a previously-identified QSO. The PMs are colored by whether \gaia\@-measured PMs and parallaxes were available as priors. The PM uncertainties are such that all of these PM measurements are consistent with stationary.}
\label{fig:QSO_vpd}
\end{center}
\end{figure}

\begin{figure}[ht]
\begin{center}
\includegraphics[width=\linewidth]{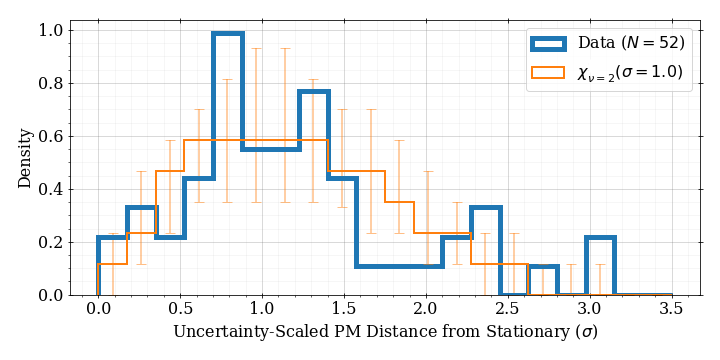}
\caption{QSO PM distance from stationary, scaled by their uncertainties. The orange histogram shows the expectation if the uncertainties explain all the non-zero sizes of the PMs, with errorbars giving the 68\% region on the density in each bin for a sample size of $N=52$; the two distributions agree quite well by eye, suggesting that the posterior PM uncertainties are reasonable and trustworthy.}
\label{fig:QSO_PM_hist}
\end{center}

\end{figure}
\begin{figure*}[ht]
\begin{center}
\includegraphics[width=\linewidth]{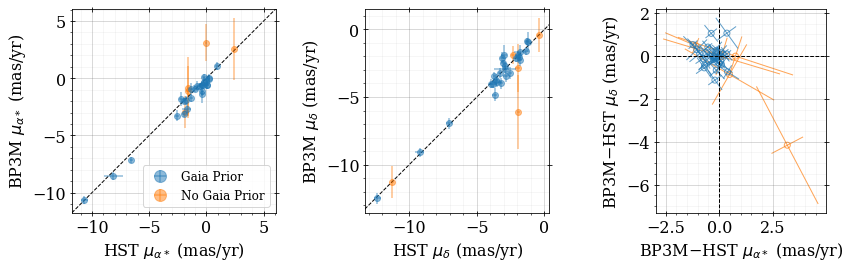}
\caption{Posterior \bpm\@ PMs for 31 sources in the COSMOS field that are shared with the HALO7D survey. The ``HST'' designation in the axis labels refer to the PMs of \citet{Cunningham_2019b}, which use \hst\@ images alone to measure PMs.}
\label{fig:COSMOS_H7D_comp}
\end{center}
\end{figure*}

As an additional method of validating the \bpm\@ posterior PMs and their uncertainties, we cross-match the $\sim2000$ unique COSMOS sources mentioned in Section~\ref{sec:gaiahub} with the MILLIQUAS QSO catalog\footnote{\url{https://heasarc.gsfc.nasa.gov/W3Browse/all/milliquas.html}} \citep{Flesch_2023}. There are 52 sources in the QSO list that have COSMOS sources within $10~\mathrm{mas}$, so we run the \hst\@ images that contain these sources through the \bpm\@ pipeline, with the resulting posterior PMs shown in Figure \ref{fig:QSO_vpd}. As expected for extremely distant sources, the PMs are closely distributed around the origin. 

Next, we use the PM uncertainties to measure the 2D distances of the PMs from stationary, and then compare these measurements to the expected distribution; the result of this process is shown in Figure \ref{fig:QSO_PM_hist}. The expected distribution (orange histogram) for an equivalent sample size agrees quite well with the measured one (blue histogram). This is additional and external evidence that the \bpm\@ posterior PM distributions are reasonable and trustworthy. 

\subsection{Comparison with \hst-measured PMs}\label{ssec:halo7d_comp}

Like in Section~\ref{ssec:QSO_comp}, we identify sources in our COSMOS sample that are shared with an external survey; in this case, we cross-match with the COSMOS sample of the HALO7D survey \citep{Cunningham_2019a,Cunningham_2019b,McKinnon_2023}. The HALO7D PMs are described in \citet{Cunningham_2019b}; to summarize, the PMs are measured using multiple epochs of \hst\@ imaging with the absolute reference frame being defined by registration with background galaxies. These \hst+\hst\@ PMs are therefore a result of traditional PM measurement techniques. 

We analyse the \hst\@ images in the multi-epoch region of COSMOS that have \gaia\@ sources in common with the HALO7D survey, finding 31 cross-matches. In Figure~\ref{fig:COSMOS_H7D_comp}, we see that the \bpm\@ PMs have strong agreement with their \hst-only counterparts. As in the other comparisons we have presented, the distribution of 2D differences between the PM measurements agree statistically with our expectations, providing the final piece of validation for the \bpm\@ outputs. 

\subsection{Proper Motions in Sparse Fields: COSMOS} \label{sec:cosmos_comp}

\begin{figure}[ht]
\begin{center}
\includegraphics[width=\linewidth]{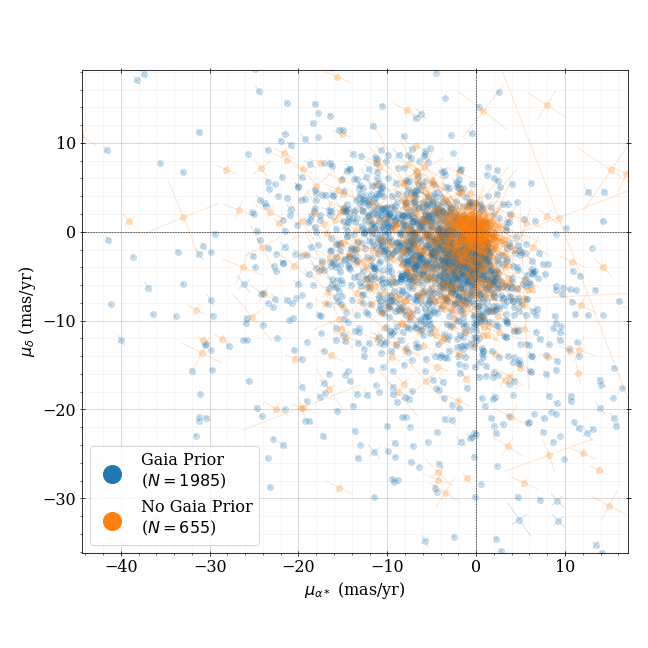}
\caption{Posterior \bpm\@ PMs for the 2640 unique sources in the COSMOS field (i.e. same sources referred to in Figure~\ref{fig:COSMOS_nstar_per_image}). While each source may fall in multiple HST images, we only show the PMs measured from analysing a single HST image with Gaia. The PMs are colored by whether \gaia\@-measured PMs and parallaxes were available as priors. The PM uncertainties of the sources with \gaia\@ priors (blue points) are, on average, much smaller than those without \gaia\@ priors (orange points).}
\label{fig:COSMOS_vpd}
\end{center}
\end{figure}

\begin{figure*}[ht]
\begin{center}
\includegraphics[width=\linewidth,height=8in,keepaspectratio, trim={0cm 0cm 0cm 0cm},clip]{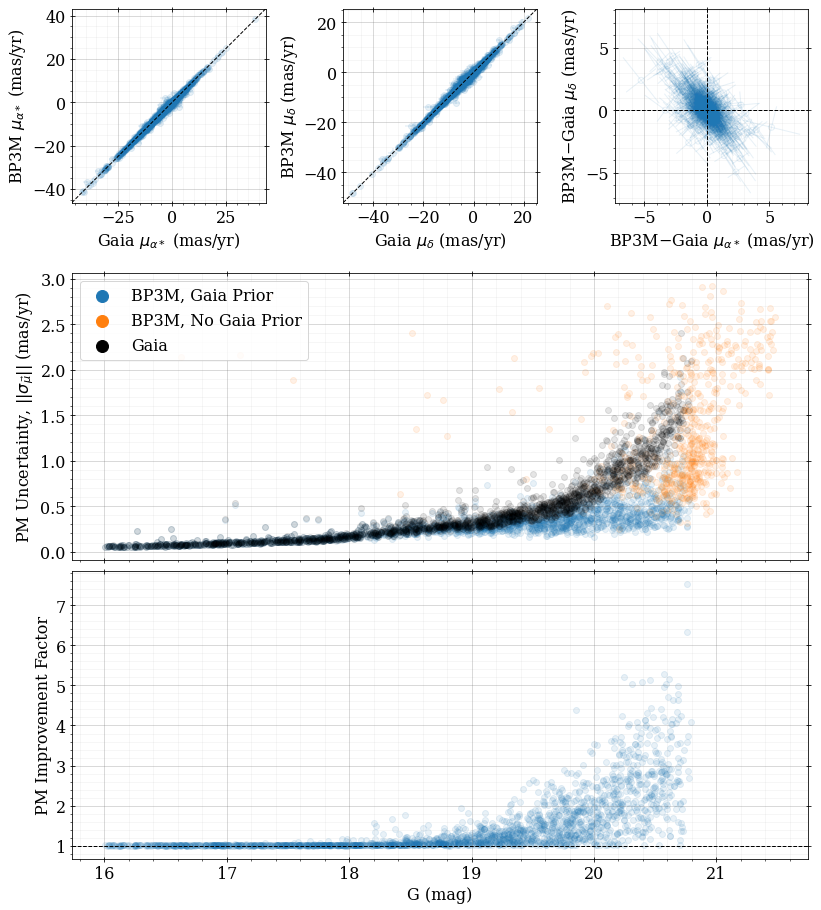}
\caption{Comparison of posterior PMs and corresponding uncertainties between \gaia\@ and \bpm\@ for the same COSMOS sources as Figure \ref{fig:COSMOS_vpd}. The different PM values agree with each other within their uncertainties across a range of PM sizes. In the upper right panel, the errorbars come from the \gaia\@-measured PM covariance matrices alone instead of a combination of \bpm\@ and \gaia\@ covariances because the \bpm\@ results depend on the \gaia\@ covariances. For many of the bright sources ($G<19~\mathrm{mag}$), their \hst\@ detection has a large position uncertainty, which leads to no improvement over the Gaia-measured PMs. For fainter targets ($G>19~\mathrm{mag}$), where the \gaia\@ PM uncertainty increases and the \hst\@ positions are better measured, there is a significant improvement in the PM uncertainty. Approximately 25\% of the COSMOS sources have no \gaia-measured PM, and these sources have a median PM uncertainty of $\sim1.16~\maspyr$.}
\label{fig:COSMOS_comp_with_Gaia}
\end{center}
\end{figure*}

Using the \hst\@ images of the COSMOS field discussed in Section~\ref{ssec:COSMOS_testbed}, we run \bpm\@ on all images within $0.5^\circ$ of the field's center. This includes all regions of COSMOS with multi-epoch \hst\@ imaging (near the center), as well as single-epoch-only \hst\@ regions. The posterior PMs we measure for the 2640 unique sources that are found in common between the \hst\@ images and \gaia\@ are shown in Figure~\ref{fig:COSMOS_vpd}, where the sources are colored by whether or not they have \gaia\@-measured PMs and parallaxes. As a quick sanity check, this PM distribution visually agrees with our expectations based on the synthetic COSMOS data created using previously-measured velocity distributions for the MW thick disk and stellar halo\footnote{See Figure~\ref{fig:synthetic_COSMOS_vpd} in Appendix~\ref{sec:generating_synthetic_data}, but note that our ``visual comparison'' here means looking at the range and spread of the PMs, not comparing the different colored populations in each figure.}. There may be, however, more \bpm\@ PMs clustered at the origin than originally expected (i.e. the orange, ``No Gaia Prior'' points), but many of these sources may truly be background stationary sources, which were not included as a component of the synthetic data. 

While each source may appear in multiple \hst\@ frames, we only show the results from analysing each \hst\@ image separately, and the displayed PM is from the posterior distribution with the smallest uncertainty. Currently, it is very computationally expensive to analyse multiple \hst\@ concurrently, making the complete simultaneous analysis of COSMOS \hst\@ images intractable. Fortunately, as we determined from our dSph test data, analyses of single \hst\@ images do not show noticeable systematics in \bpm\@ PMs as a result of the \hst\@ position uncertainty estimates, so the PMs we present for the COSMOS sample are robust. After increasing computation efficiency and \hst\@ position uncertainty estimates in planned updates to the pipeline, we will be able to present a complete COSMOS analysis in future work. 

We also compare the \bpm\@ PMs and uncertainties to the corresponding \gaia\@ PMs in Figure~\ref{fig:COSMOS_comp_with_Gaia}, where we again see excellent agreement between the two samples (top panels). Because the sources in this figure include the same sources shown in Figure~\ref{fig:COSMOS_gaiahub_comp}, we can also compare the sparse-field PM performance between the \gaiahub\@ and \bpm\@ pipelines. In general, \bpm\@ returns much closer agreement with the \gaia\@ PMs, as is expected because it uses those measurements as prior constraints. In addition, \bpm\@ is able to measure consistent PMs for all 2640 \gaia\@ sources in the COSMOS images, which is a large increase from the 1659 PMs catalog of \gaiahub\@. These results suggest that \bpm\@'s approach has successfully addressed the issues producing the outliers of Figure~\ref{fig:COSMOS_comp_with_Gaia}, indicating it is a valid method of extracting useful and trustworthy PMs in sparse fields. 

The middle and bottom panels of Figure~\ref{fig:COSMOS_comp_with_Gaia} show the improvement on the PM uncertainty that \bpm\@ measures by combining \hst\@ and \gaia\@, especially for the $\sim25\%$ of sources that had no \gaia-measured PMs; the median PM uncertainty for sources without \gaia\@ priors is found to be $\sim1.16~\maspyr$. While many of the brightest magnitude sources -- those most likely to be at or near saturation in the \hst\@ images -- show no improvement on \gaia's PMs, the faintest magnitude sources ($20.25<G<20.75~\mathrm{mag}$) have a median PM improvement factor of 2.6. As mentioned for the small-source-count case in Figure~\ref{fig:fake_pm_comparison_n010}, the less extreme improvements in PM we see in sparse fields are likely the result of larger uncertainty in the transformation parameters, which propagates to the PM uncertainties of all sources in an image. 

The \bpm\@ transformation parameters could likely be pinned down much more precisely by identifying stationary sources in each image. For example, by finding the centroids in each \hst\@ image that are clearly associated with quasars and background galaxies, then updating the priors in parallax and PM to a narrow distribution around $\vec 0$, the transformation solution can be improved by these anchor points. This would reduce the posterior width of the transformation parameter distributions for each image, especially for particularly sparse images, which would also reduce the posterior widths in the PMs for all of the sources in an \hst\@ image. Exploring the impact that known stationary targets can have on the \bpm\@ PMs will be the focus of future work. 

The median PM uncertainties in the $20<G<21.5~\mathrm{mag}$ range are summarized for the COSMOS sample in Table~\ref{tab:vsky_uncertainty_vs_distance}. 
First, the \bpm\@ sample that has \gaia\@ parallax and PM priors has a significant increase in precision compared to \gaia\@ alone. Second, the \gaia\@ COSMOS sample ($20<G<20.75~\mathrm{mag}$) has a similar median PM uncertainty to the \bpm\@ sample without \gaia\@ priors ($20.75<G<21.5~\mathrm{mag}$), but the latter contains significantly more stars. Effectively, \bpm\@ substantially increases the number of stars in a single field with \gaia\@-quality PMs at the faintest magnitudes. 

\begin{table}[t]
    \centering
    \caption{Median uncertainty in PM for the COSMOS stars in \gaia\@, \bpm\@ with \gaia\@ priors, and \bpm\@ without \gaia\@ priors in the $20<G<21.5~\mathrm{mag}$ range. While the median uncertainties in the \gaia\@ sample are smaller than the medians from the \bpm\@ without \gaia\@ priors sample, the former has significantly fewer stars than the latter and does not reach as faint. As shown in the middle panel of Figure~\ref{fig:COSMOS_comp_with_Gaia}, the \gaia\@ COSMOS data extend only to $G\sim20.75~\mathrm{mag}$, while the \bpm\@ sample extends to $G\sim21.5~\mathrm{mag}$.}
    \label{tab:vsky_uncertainty_vs_distance}
    \begin{tabular}{ccccccc}
\hline \hline
 & Number of & Median $||\sigma_{\vec\mu_i}||$ \\
Sample & Sources & (mas/yr) \\ \hline
\gaia\@ & 443 & 1.03 \\
\bpm, Gaia Prior & 443 & 0.44 \\
\bpm, No Gaia Prior & 632 & 1.13 \\    
\hline
\end{tabular}
\end{table}

A typical MW stellar halo isochrone (i.e. $\age=12~\mathrm{Gyr}$ and $\feh=-1.2~\mathrm{dex}$, see Appendix~\ref{sec:generating_synthetic_data}) has a red giant branch (RGB) base around an absolute magnitude of $M_G\approx3~\mathrm{mag}$. For stars at the base of the RGB with apparent magnitudes in the $20<G<20.75~\mathrm{mag}$ range, this translates to a distance range of $\sim25-35~\mathrm{kpc}$. Using the median PM uncertainties in Table~\ref{tab:vsky_uncertainty_vs_distance}, this would imply an uncertainty in velocity on the sky in the range of $\sim123-173~\kmps$ for the \gaia-alone sample versus a $\sim52-74~\kmps$ range for the \bpm\@ sample with \gaia\@ priors. For the \bpm\@ sample without \gaia\@ priors, the apparent magnitude range of $20.75<G<21.5~\mathrm{mag}$ translates to a distance range of $\sim35-50~\mathrm{kpc}$, and results in a sky velocity uncertainty range of $\sim190-268~\kmps$ when using the median PM uncertainty for this subsample. 

Like with the dwarf spheroidal data, many of these COSMOS sources appear in multiple \hst\@ images. As a result, the \bpm\@ PM uncertainties presented in this section are significantly larger than the possible PM uncertainty that could be achieved by analysing the images concurrently, which will be the focus on future work. This may also likely yield improvements on the parallax precision, especially for sources in the multi-epoch \hst\@ regions of the field. Finally, there are many \hst\@ sources that are too faint to have \gaia\@ counterparts, but they nonetheless appear in multiple \hst\@ frames. For these sources, future versions of \bpm\@ can use the \hst-to-\gaia\@ transformation parameters for each image to identify the positions of these sources in a global reference frame as a function of time, and thus determine their best fit PMs and parallaxes. 

In tests of a handful of COSMOS images where we analyse all overlapping \hst\@ concurrently, we measure a median PM uncertainty of $0.064~\maspyr$ for $20<G<20.75$~mag (sources with \gaia\@-measured PMs) and a median PM uncertainty of $0.25~\maspyr$ for $20.75<G<21.5$~mag (sources with \gaia\@-measured PMs). In future work, we plan to use the extreme precision PMs to investigate substructure in the MW stellar halo at large distances ($D>40$~kpc). Other future applications including measuring internal kinematics of perturbed stellar streams -- whose stars are often in sparse fields that are analogous to COSMOS, and where the 3D velocity error budget is dominated by PM terms \citep[e.g., GD-1][]{Price_Whelan_2018,Malhan_2019} -- to constrain the dark matter subhalo mass function. 

\section{Summary} \label{sec:summary}

We have presented the statistics that describe, in general, how positions of sources from two or more sets of images (from any telescope or instrument) should be combined in a Bayesian hierarchical framework to measure transformation parameters between the images, as well as distributions on the positions, parallaxes, and PMs for the individual sources. We use these statistics to created \bpm\@ -- a pipeline that builds off of \gaiahub\@ -- to combine archival \hst\@ and \gaia\@ data to measure reliable positions, parallaxes, and PMs in the absolute reference frame of \gaia, even in sparse fields (e.g. $N_* < 10$ per \hst\@ image). We have tested this pipeline using a combination of synthetic \hst\@ images, nearby dwarf galaxies, a catalog of QSOs, and sparse MW stellar halo fields. In all cases, we find that the \bpm\@ PMs are both accurate and precise, especially for magnitudes where \gaia\@ has large PM uncertainties ($19<G<21~\mathrm{mag}$) or no PM measurements at all ($21<G<21.5~\mathrm{mag}$).

Our key results include:
\begin{enumerate}
    \item When trying to maximize parallax precision, the best observing strategy we have found is to alternate the observation times between the two semi-major axis extremes of the ellipse created by the parallax motion. The improvement to the parallax and position precision come at no cost to the PM uncertainty (Section~\ref{ssec:recover_parallaxes}, Figures~\ref{fig:synthetic_HST_uncertainties});
    \item For time baselines of 10 to 13~years, the \bpm\@ pipeline is able to measure PMs of stars in nearby dwarf spheroidals with uncertainties that are, on average, 8 to 13 times more precise than \gaia\@ alone for $20.5<G<21~\mathrm{mag}$ (Section~\ref{ssec:dSph_comp}, Figures~\ref{fig:fornax_pm_comparison} and \ref{fig:draco_pm_comparison});
    \item For the sparse fields of COSMOS (median 9 sources shared with \gaia\@ per \hst\@ image), we measure a median PM improvement factor of 2.6 for sources with \gaia\@-measured PMs, and find a median PM uncertainty of $1.16~\maspyr$ for the $\sim25\%$ of sources without \gaia\@-measured PMs (Section~\ref{sec:cosmos_comp}, Figure~\ref{fig:COSMOS_comp_with_Gaia}).
\end{enumerate}

\begin{acknowledgments}

KM, CMR, PG, and MA were funded by NASA through grants associated with HST Archival Program AR-16625 awarded by the Space Telescope Science Institute, which is operated by the Association of Universities for Research in Astronomy, Inc., for NASA, under contract NAS5-26555. AdP acknowledges the financial support from the European Union - NextGenerationEU and the Spanish Ministry of Science and Innovation through the Recovery and Resilience Facility project J-CAVA and the project PID2021-124918NB-C41.

KM thanks the individuals and teams of researchers whose decades of work to characterize the optics of \hst\@ that have made this work possible. KM also thanks Melodie Kao for code that helped implement the motion-from-parallax calculations. 

\end{acknowledgments}

This work was carried out in collaboration with the  HSTPROMO (High-resolution Space Telescope PROper MOtion) Collaboration\footnote{\url{https://www.stsci.edu/~marel/hstpromo.html}}, a set of projects aimed at improving our dynamical understanding of stars, clusters, and galaxies in the nearby Universe through measurement and interpretation of proper motions from HST, Gaia, and other space observatories. We thank the collaboration members for sharing their ideas and software.

\vspace{5mm}
\facilities{\hst\@, \gaia\@}

\software{\texttt{Astropy} \citep{astropy_2013,astropy_2018,astropy_2022}, \texttt{corner} \citep{corner_citation}, \texttt{emcee} \citep{emcee_citation}, \gaiahub\@ \citep{delPino_2022}, \texttt{IPython} \citep{ipython_citation}, \texttt{jupyter} \citep{jupyter_citation}, \texttt{matplotlib}  \citep{matplotlib_citation}, \texttt{numpy} \citep{numpy_citation}, \texttt{pandas} \citep{pandas_citation_2010,pandas_citation_2020}, \texttt{scipy} \citep{scipy_citation}}

\bibliography{main_ARXIV}{}
\bibliographystyle{aasjournal}

\appendix
\onecolumngrid

\section{Motion Statistics} \label{sec:motion_statistics}

Here, we present the posterior conditional distributions in PM, parallax, and position for a given source. 

Because of it's lack of coefficients, it is easiest to solve for the posterior conditional on $\vec{\Delta \theta_i}$ first:
\begin{equation}
    \begin{split}
        \left(\vec{\Delta \theta_i} | \vec\mu_i, \mathrm{plx}_i,\dots \right) \sim \mathcal{N}&\left( \pmb \Sigma_{\theta,i} = \left[\pmb V^{-1}_{\theta,i}+ \sum_{j=1}^{n_{im}} \pmb J_{ij}^{T} \cdot \pmb V^{-1}_{d,ij} \cdot \pmb J_{ij}\right]^{-1}, \right. \\ 
        &\hspace{0.5cm} \left. \vec \mu_{\theta,i} = \pmb \Sigma_{\theta,i}\cdot\left[\pmb V^{-1}_{\theta,i}\cdot \vec{\Delta \theta}'_{i} + \sum_{j=1}^{n_{im}} \pmb J_{ij}^{T} \cdot \pmb V^{-1}_{d,ij} \cdot \pmb J_{ij} \cdot \left(\pmb{\mathrm{plx}_i} \cdot \vec{\Delta \mathrm{plx}_{ij}}+ \pmb{\Delta t_{j}}\cdot \vec\mu_i  - \pmb J^{-1}_{ij}\cdot \vec{\Delta d_{ij}}\right) \right]\right)
    \end{split}
\end{equation}
where we use bold-faced versions of scalars to represent the corresponding identity matrix version of that number, such as
$$\pmb{\mathrm{plx}}_i = \mathrm{plx}_i \cdot \pmb I_{2\times2}.$$ We have also included a term for the \gaia\@-measured position offset vector $\vec{\Delta \theta}_i'$ to be as general as possible, but this vector is just $\vec0$ because \gaia\@ has no measured offset from the positions it reports. 

Next, the posterior conditional on $\vec \mu_i$ is given by:
\begin{equation}
    \begin{split}
        (\vec \mu_i |\mathrm{plx}_i \dots) &\sim \mathcal{N}\left(
            \pmb\Sigma_{\mu,i} = \left[\pmb V_{\mu,i}^{-1} + \pmb{\hat V_{\mu}}^{-1} +\pmb \Sigma_{\mu,\theta,i}^{-1}+\sum_{j=1}^{n_{im}} \pmb \Sigma_{\mu,d,ij}^{-1} \right]^{-1}, \right.\\
            &\hspace{1.23cm}\left.\vec \mu_{\mu,i} = \pmb\Sigma_{\mu,i} \cdot \left[ \pmb V_{\mu,i}^{-1}\cdot \vec \mu_i' + \pmb{V_{\hat \mu}}^{-1}\cdot  \hat \mu+\pmb \Sigma_{\mu,\theta,i}^{-1} \cdot \vec \mu_{\mu,\theta,i} +\sum_{j=1}^{n_{im}} \pmb \Sigma_{\mu,d,ij}^{-1} \cdot \pmb C_{\mu,ij}^{-1} \cdot \vec D_{\mu,ij}\right] \right)
    \end{split}
\end{equation}
where
\begin{equation}
\pmb A_{\mu,i} = \left[ \pmb \Sigma_{\theta,i} \cdot \sum_{j=1}^{n_{im}} \pmb J_{ij}^T \cdot \pmb V_{d,ij}^{-1}\cdot \pmb J_{ij} \cdot \Delta t_{j} \right],
\end{equation}
\begin{equation}
\vec B_{\mu,i} = \pmb \Sigma_{\theta,i} \cdot \left[\sum_{j=1}^{n_{im}} \pmb J_{ij}^T \cdot \pmb V_{d,ij}^{-1}\cdot \pmb J_{ij} \cdot \pmb J_{ij}^{-1}\cdot \vec{\Delta d_{ij}} -\pmb V_{\theta,i}^{-1}\cdot \vec{\Delta \theta_i'}  - \mathrm{plx}_i  \cdot \sum_{j=1}^{n_{im}} \pmb J_{ij}^T \cdot \pmb V_{d,ij}^{-1}\cdot \pmb J_{ij} \cdot \vec{\Delta \mathrm{plx}_{ij}}\right],
\end{equation}
\begin{equation}
\pmb C_{\mu,ij} = \pmb{\Delta t_j} - \pmb A_{\mu,i},
\end{equation}
\begin{equation}
\vec{D_{\mu,ij}} = \pmb J_{ij}^{-1} \cdot \vec{\Delta d_{ij}} - \vec B_{\mu,i} - \pmb{\mathrm{plx}_i} \cdot \vec{\Delta \mathrm{plx}_{ij}}.
\end{equation}

Finally, the posterior conditional on $\mathrm{plx}_i$ is given by:
\begin{equation}
    \begin{split}
        (\mathrm{plx}_i | \dots) \sim & \mathcal{N}\left(\mathrm{plx}_i | \hat{\sigma_{\mathrm{plx},i}}^2 = \left[ \hat\sigma_{\mathrm{plx}}^{-2}+\sigma^{-2}_{\mathrm{plx},i} + \sigma^{-2}_{\mathrm{plx},\theta,i} + \sigma^{-2}_{\mathrm{plx},\mu,i} + \sigma^{-2}_{\mathrm{plx},\hat\mu,i} + \sum_{j=1}^{n_{im}} \sigma^{-2}_{\mathrm{plx},d,ij}\right]^{-1}, \right.\\
            & \hspace{1.5cm}\left. \hat{\mu_{\mathrm{plx},i}} = \hat{\sigma_{\mathrm{plx},i}}^2 \cdot \left[ \hat\sigma_{\mathrm{plx}}^{-2} \cdot \hat{\mathrm{plx}}+ \sigma^{-2}_{\mathrm{plx},i}\cdot \mathrm{plx}_i' + \sigma^{-2}_{\mathrm{plx},\theta,i}\cdot\mu_{\mathrm{plx},\theta,i} + \sigma^{-2}_{\mathrm{plx},\mu,i}\cdot\mu_{\mathrm{plx},\mu,i} \right. \right.\\
            & \hspace{8cm} \left. \left. + \sigma^{-2}_{\mathrm{plx},\hat\mu,i}\cdot\mu_{\mathrm{plx},\hat\mu,i} + \sum_{j=1}^{n_{im}} \sigma^{-2}_{\mathrm{plx},d,ij} \cdot \mu_{\mathrm{plx},d,ij}\right] \right)
        \end{split}
\end{equation}
where
\begin{equation}
\vec{A_{\mathrm{plx},\mu,i}} = -\pmb \Sigma_{\theta,i}   \cdot \sum_{j=1}^{n_{im}} \pmb J_{ij}^T \cdot \pmb V_{d,ij}^{-1}\cdot \pmb J_{ij} \cdot \vec{\Delta \mathrm{plx}_{ij}},
\end{equation}
\begin{equation}
\vec{B_{\mathrm{plx},\mu,i}} = \pmb \Sigma_{\theta,i} \cdot \left[\pmb V_{\theta,i}^{-1}\cdot \vec{\Delta \theta_i'} -  \sum_{j=1}^{n_{im}} \pmb J_{ij}^T \cdot \pmb V_{d,ij}^{-1}\cdot \pmb J_{ij} \cdot \pmb J_{ij}^{-1}\cdot \vec{\Delta d_{ij}} \right],
\end{equation}
\begin{equation}
    \vec C_{\mathrm{plx},\mu,i} = \pmb\Sigma_{\mu,i} \cdot \left[\pmb \Sigma_{\mu,\theta,i}^{-1} \cdot \pmb A_{\mu,i}^{-1} \cdot \vec{A_{\mathrm{plx},\mu,i}} -\sum_{j=1}^{n_{im}} \pmb \Sigma_{\mu,d,ij}^{-1} \cdot \pmb C_{\mu,ij}^{-1} \cdot \left(  \vec{A_{\mathrm{plx},\mu,i}} + \vec{\Delta \mathrm{plx}_{ij}}   \right)\right],
\end{equation}
\begin{equation}
    \begin{split}
        \vec D_{\mathrm{plx},\mu,i} =& -\pmb\Sigma_{\mu,i} \cdot \left[\pmb V_{\mu,i}^{-1}\cdot \vec \mu_i' + \pmb{V_{\hat \mu}}^{-1}\cdot  \hat \mu+ \pmb \Sigma_{\mu,\theta,i}^{-1} \cdot \pmb A_{\mu,i}^{-1}\cdot  \left( \vec{\Delta \theta_i}' - \vec{B_{\mathrm{plx},\mu,i}}\right)\right.\\
        &\hspace{5cm} \left. +\sum_{j=1}^{n_{im}} \pmb \Sigma_{\mu,d,ij}^{-1} \cdot \pmb C_{\mu,ij}^{-1} \cdot \left( \vec{B_{\mathrm{plx},\mu,i}} + \pmb J_{ij}^{-1} \cdot \vec{\Delta d_{ij}}  \right)\right],
    \end{split}
\end{equation}
\begin{equation}
\vec G_{\mathrm{plx},d,ij} = \left[\vec{\Delta \mathrm{plx}_{ij}} +\Delta t_{j}\cdot \vec{C_{\mathrm{plx},\mu,i}} -\vec{E_{\mathrm{plx},\theta,i}}\right],
\end{equation}
\begin{equation}
\vec H_{\mathrm{plx},d,ij} = \left[\Delta t_{j}\cdot \vec{D_{\mathrm{plx},\mu,i}} - \vec{F_{\mathrm{plx},\theta,i}} + \pmb J_{ij}^{-1}\cdot \vec{\Delta d_{ij}}\right]. \end{equation}

Because these three posterior conditional distributions are all Gaussian, we are able to quickly sample PMs, parallaxes, and position offsets for source $i$ given a set of transformation parameters. \vspace{0.5cm}

\begin{table*}[b]
\centering
\caption{Distributions used to generate synthetic data of halo and thick MW stars. $\mathcal{SKN}$ indicates a skew-normal distribution.}
\begin{tabular}{c|c|c}
\hline \hline
Component & Distribution                                   & Functional Form \\ \hline
Halo &  $p(\feh)$ & $0.406 \cdot\mathcal{SKN}(\mu=-1.35~\mathrm{dex},\sigma=0.5~\mathrm{dex},a=5)\hspace{3cm}$ \\ 
 &   & $\hspace{3cm}+0.594\cdot\mathcal{SKN}(\mu=-0.9~\mathrm{dex},\sigma=0.7~\mathrm{dex},a=-3)$ \\ 
 & $p(\mathrm{age})$            & $\mathcal{N}(\mu=12~\mathrm{Gyr},\sigma=2~\mathrm{Gyr})$          \\
 & $p(\mathrm{mass} | \feh, \mathrm{age})$            & \citet{Kroupa_2001} IMF,   $\propto k \left(\frac{\mathrm{mass}}{M_\odot}\right)^{-\alpha}$ with
        $\begin{cases}
            k=25,~\alpha = 0.3, & \mathrm{mass} < 0.08 M_\odot \\
            k=2,~\alpha = 1.3, & \mathrm{mass} < 0.5 M_\odot \\
            k=1,~\alpha = 2.3, & \mathrm{mass} > 0.5 M_\odot
        \end{cases}$ \\
 & $p(\mathrm{distance~moludus})$  & $\propto D^3\left(\frac{R_q}{27 \mathrm{kpc}}\right)^{-\alpha}$ with
        $\begin{cases}
            \alpha = 2.3, & R_q < 27\, \mathrm{kpc} \\
            \alpha =4.6, & R_q \geq 27\, \mathrm{kpc} 
        \end{cases}$ \\  
 & $p(v_r)$ &      $\mathcal{N}(\mu=0~\kmps, \sigma=130~\kmps) $    \\
 &  $p(v_\phi)$ &   $\mathcal{N}(\mu=0~\kmps, \sigma=70~\kmps)$ \\
 & $p(v_\theta)$  &   $ \mathcal{N}(\mu=0~\kmps, \sigma=70~\kmps)$    \\ \hline 
Thick Disk &  $p(\feh)$ & $\frac{1}{6}\mathcal{SKN}(\mu=-1.05~\mathrm{dex},\sigma=0.6~\mathrm{dex},a=-5)\hspace{3cm}$ \\ 
 &   & $\hspace{3cm}+\frac{5}{6}\mathcal{N}(\mu=-0.54~\mathrm{dex},\sigma=0.3~\mathrm{dex})$ \\ 
 & $p(\age)$ & $ \mathcal{N}(\mu=10~\mathrm{Gyr},\sigma=2~\mathrm{Gyr})$ \\
 & $p(\mathrm{mass} | \feh, \mathrm{age})$            & \citet{Kroupa_2001} IMF      \\
 & $p(\mathrm{distance~moludus})$            &  $\propto D^3 \exp\left(-\frac{R_D}{3~\mathrm{kpc}}-\frac{z}{1~\mathrm{kpc}}\right)$, where $R_D^2 = x^2+y^2$ \\
 &  $p(v_{R_D})$ &  $ \mathcal{N}(\mu=0~\kmps, \sigma=45~\kmps) $         \\
 &  $p(v_z)$ &  $ \mathcal{N}(\mu=0~\kmps, \sigma=20~\kmps) $ \\
 & $p(v_T)$  &  $ \mathcal{SKN}(\mu = 242~\kmps, \sigma=46.2~\kmps,a=-2) $    \\ 
 \hline 
\end{tabular}
\label{tab:disk_halo_distributions}
\end{table*}

\twocolumngrid
\section{Generating Synthetic, COSMOS-like Data} \label{sec:generating_synthetic_data}

\begin{figure*}[t]
\begin{center}
\begin{minipage}[c]{0.49\linewidth}
\includegraphics[width=\linewidth]{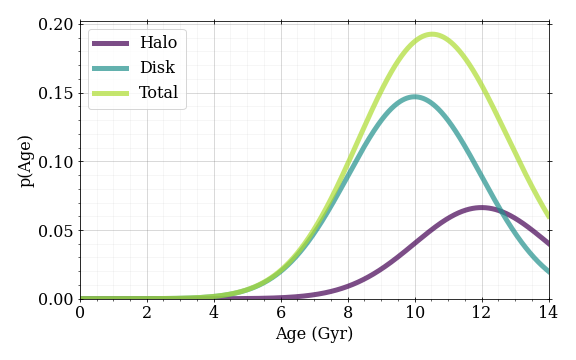}
\end{minipage} \hfill
\begin{minipage}[c]{0.49\linewidth}
\includegraphics[width=\linewidth]{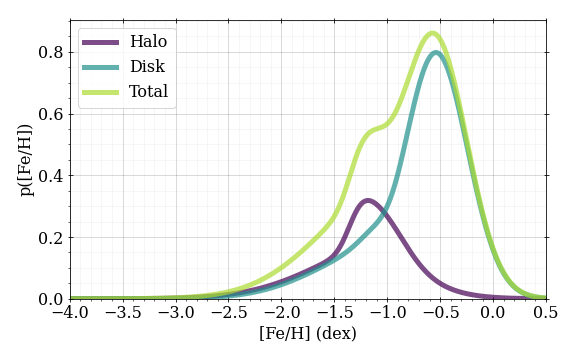}
\end{minipage}
\caption{Distributions on age (left panel) and $\feh$ (right panel) for the MW halo (purple) and thick disk (turquoise) populations used to create synthetic stars. The halo population makes up 28\% of the total population (lime green), while the thick disk is the other 72\%.}
\label{fig:synthetic_COSMOS_MDF_and_ADF}
\end{center}
\end{figure*}

\begin{figure}[h]
\begin{center}
\includegraphics[width=\linewidth,trim={0cm 0.5cm 1cm 2.5cm},clip]{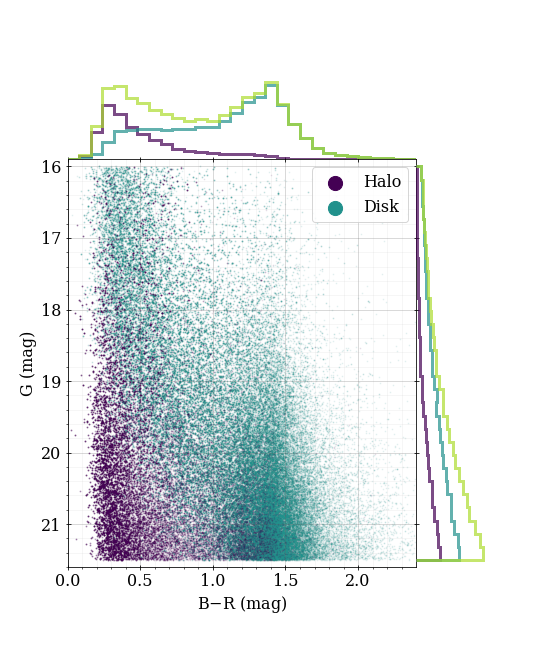}
\caption{Color-magnitude diagram of synthetic COSMOS-like stars, colored by whether the stars belong to the MW halo (purple) or thick disk (turquoise). The histograms along the right and top edges include the total population (lime green).}
\label{fig:synthetic_COSMOS_CMD}
\end{center}
\end{figure}

\begin{figure}[h]
\begin{center}
\includegraphics[width=\linewidth,trim={0cm 0.5cm 0cm 0.5cm},clip]{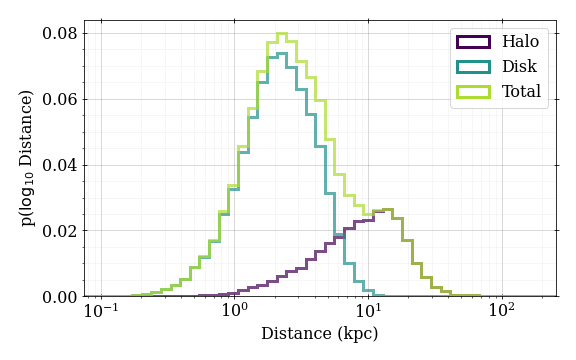}
\caption{Distance distributions of synthetic COSMOS-like stars, colored by whether the stars belong to the MW halo (purple) or thick disk (turquoise).}
\label{fig:synthetic_COSMOS_dist_hist}
\end{center}
\end{figure}

\begin{figure}[h]
\begin{center}
\includegraphics[width=\linewidth,trim={0cm 1.5cm 0cm 3.cm},clip]{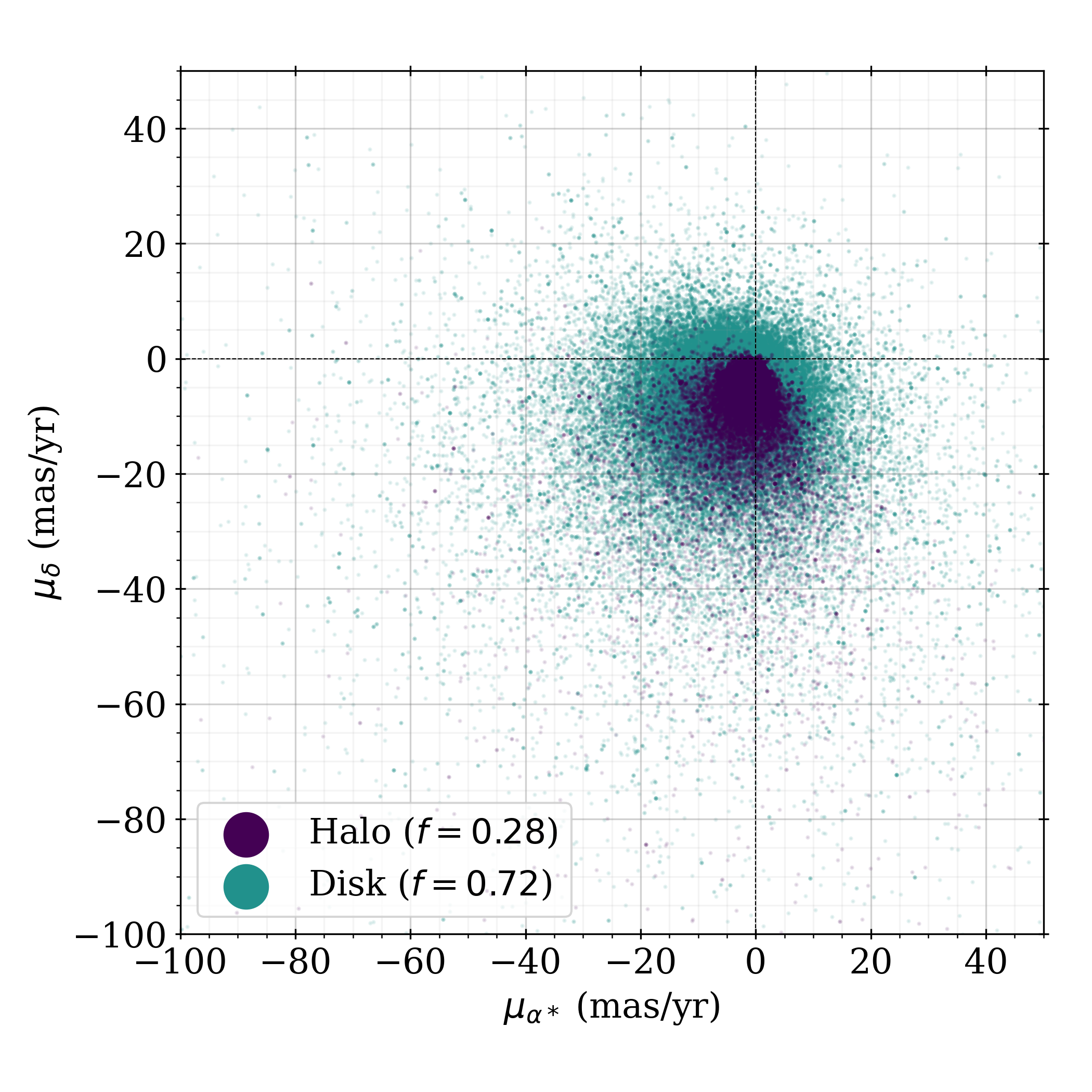}
\caption{PMs of synthetic COSMOS-like stars, colored by whether the stars belong to the MW halo (purple) or thick disk (turquoise).}
\label{fig:synthetic_COSMOS_vpd}
\end{center}
\end{figure}

\begin{figure*}[t]
\begin{center}
\begin{minipage}[c]{0.32\linewidth}
\includegraphics[width=\linewidth]{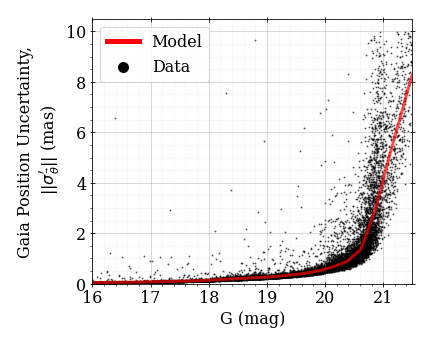}
\end{minipage} \hfill
\begin{minipage}[c]{0.32\linewidth}
\includegraphics[width=\linewidth]{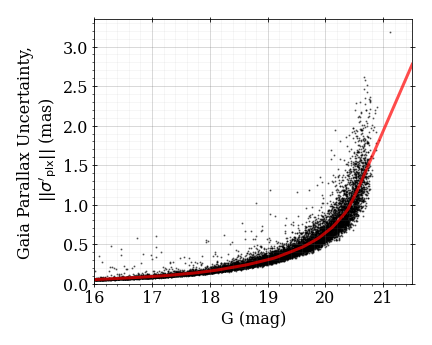}
\end{minipage} \hfill
\begin{minipage}[c]{0.32\linewidth}
\includegraphics[width=\linewidth]{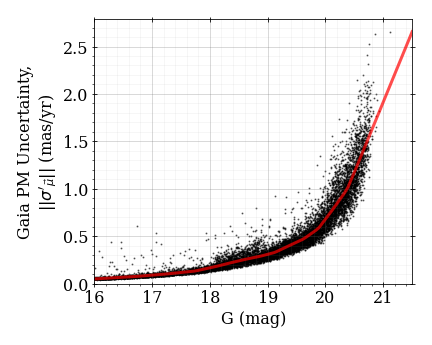}
\end{minipage}
\caption{\gaia\@ Position, parallax, and PM uncertainties as a function of magnitude from all \gaia\@ sources within 1~degree of the COSMOS field center. The black points are the \gaia-measured values and the red lines are a median-binning of those data, with a linear extrapolation where no \gaia\@ measurements exist.}
\label{fig:synthetic_COSMOS_Gaia_uncertainties}
\end{center}
\end{figure*}

Following a similar approach to \citet{McKinnon_2023}, we define mass, age, $\feh$, and distance distributions for both a MW thick disk and a MW halo population. We also define distributions for the Galactocentric 3D velocities for the halo and thick disk. The distributions for each parameter that go into generating a synthetic star are summarized in Table~\ref{tab:disk_halo_distributions}. The particular choices for the velocity distributions come from the MW halo velocity ellipsoid measurements of \citet{Cunningham_2019b}, as do the distance modulus distributions. For the age distributions, we use an analytical approximation of the age distributions presented in \citet{Bonaca_2020}. The $\feh$ distributions are again analytical approximations of the results presented in \citet{Conroy_2019b} and \citet{Naidu_2020}. The age and $\feh$ distributions are displayed in Figure~\ref{fig:synthetic_COSMOS_MDF_and_ADF}, where the distributions have been scaled by their contribution to the total halo population. After consulting a Besan\c{c}on model \citep{Robin_2003} in a COSMOS-like field in the $16<G<21.5~\mathrm{mag}$ range, we choose to set the fraction of the number of observed stars that belong to the halo (i.e. ``halo fraction'') to 28\%.

Using these distributions and the halo fraction, we draw stellar parameters (i.e. mass, age, and $\feh$) for each synthetic star, and then interpolate to that point using the MIST isochrones\footnote{\url{https://waps.cfa.harvard.edu/MIST/index.html}} \citep{Dotter_2016,Choi_2016,Paxton_2011,Paxton_2013,Paxton_2015}. This interpolated point yields the color and absolute magnitude of each synthetic star. After we draw distances, we can measure the apparent magnitudes for each star, and then perform rejection sampling to get a sample of stars that fall within a particular range in magnitude and/or color. A color-magnitude diagram is shown in Figure~\ref{fig:synthetic_COSMOS_CMD} for the COSMOS-like stars we generate, with histograms along the top and right edges showing the distribution of the halo (purple), thick disk (turquoise), and total (lime green) populations. A histogram of the heliocentric distance to each synthetic star is given in Figure~\ref{fig:synthetic_COSMOS_dist_hist}.

Next, for each synthetic star, we draw the 3D velocity components based on whether that star belongs to the halo or the thick disk. Using the stellar position on the sky and distance allow us to transform those Galactocentric velocities into observable velocities\footnote{To transform between the observer frame and a Galactocentric one, we use $r_\odot = 8.1$~kpc, assume a circular speed of $235~{\rm km \ s}^{-1}$, and solar peculiar motion $(U,V,W) = (11.1,12.24,7.25)~{\rm km \ s}^{-1}$ \citep{Schonrich_2011}.}. This yields the PMs shown in Figure~\ref{fig:synthetic_COSMOS_vpd}.

To create outputs that the \bpm\@ pipeline expects, the next step is to create synthetic \hst\@ images and corresponding \gaia\@ measurements. For synthetic \gaia\@ uncertainties on position, parallax, and PM, we look at real \gaia-measured stars within 1~degree of the COSMOS field center; the resulting uncertainties in each dimension are shown in Figure \ref{fig:synthetic_COSMOS_Gaia_uncertainties} as a function of magnitude (black points) with a median-binned line overlaid (red). In cases where the data do not extend as faint as needed, we linearly extrapolate from the median binning. 

\begin{figure}[t]
\begin{center}
\includegraphics[width=\linewidth]{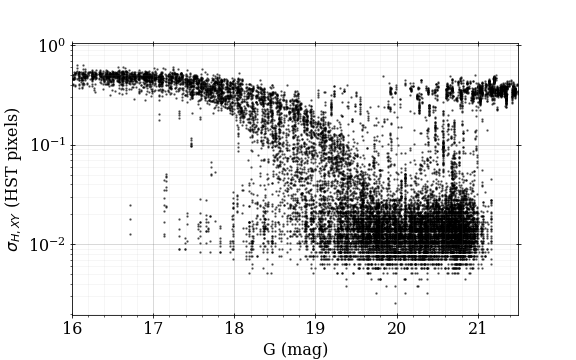}
\caption{Centroid position uncertainty in \hst\@ pixels as a function of magnitude as measured by \gaiahub\@ for the $\sim2000$ real COSMOS stars in Figure \ref{fig:COSMOS_vpd}, where the y-axis is in log scale. In general, the brighter magnitude sources have larger \hst\@ position uncertainties because they are more likely to be saturated in the COSMOS \hst\@ exposures. At the faintest magnitudes, there are some sources with large \hst\@ position uncertainties because their PSF-fitting was not as well-constrained as some other faint sources.}
\label{fig:synthetic_COSMOS_HST_uncertainty}
\end{center}
\end{figure}

We use a similar approach when it comes to modeling the uncertainties in the \hst\@ image positions. Specifically, we use the $\sim2000$ real COSMOS stars from Section~\ref{sec:cosmos_comp} to look at the \gaiahub\@-measured \hst\@ position uncertainties as a function of magnitude, and this is shown in Figure~\ref{fig:synthetic_COSMOS_HST_uncertainty}. When we go to assign \hst\@ position uncertainties to the synthetic stars, we identify the 10 nearest real COSMOS stars in $G$ magnitude and randomly select one of those stars to bequeath its position uncertainty to the synthetic one. 

With all of the true properties of the synthetic star defined, we can create \gaia-like measurements of the position, parallax, and uncertainty by applying some noise corresponding to the uncertainties we've defined. Those true positions and motions can then be played forward or backward in time until reaching the correct epoch of the synthetic \hst\@ image. After choosing the transformation parameters that map the synthetic \hst\@ image onto the synthetic \gaia\@ data, we can apply the correct \hst\@ position noise, and then save the outputs in the same \texttt{.csv} files that \bpm\@ expects as inputs. The final step of the process is to run \bpm\@ on the newly-generated synthetic \hst\@ images. 

\end{document}